\documentclass[12pt,journal,onecolumn,draftclsnofoot]{IEEEtran}%
\usepackage[dvipdfm]{graphicx}
\usepackage[usenames,dvipsnames]{color}
\usepackage{amsmath}%
\usepackage{amsfonts}%
\usepackage{amssymb}%
\usepackage{graphicx}%
\usepackage{multicol}%
\usepackage{verbatim}%
\usepackage{enumerate}%
\begin{document}

\title{Cognitive Radio Transmission Strategies for Primary Markovian Channels}

\author{
\authorblockN{Ahmed ElSamadouny, Mohammed Nafie and Ahmed Sultan}\\
\authorblockA{Wireless Intelligent Networks Center (WINC)\\
Nile University, Cairo, Egypt\\
Email: ahmed.elsamadony@nileu.edu.eg, mnafie@nileuniversity.edu.eg, salatino@stanfordalumni.org}}
\maketitle

\begin{abstract}
A fundamental problem in cognitive radio systems is that the cognitive radio is ignorant of the primary channel state and, hence, of the amount of actual harm it inflicts on the primary license holder. Sensing the primary transmitter does not help in this regard. To tackle this issue, we assume in this paper that the cognitive user can eavesdrop on the ACK/NACK Automatic Repeat reQuest (ARQ) fed back from the primary receiver to the primary transmitter. Assuming a primary channel state that follows a Markov chain, this feedback gives the cognitive radio an indication of the primary link quality. Based on the ACK/NACK received, we devise optimal transmission strategies for the cognitive radio so as to maximize a weighted sum of primary and secondary throughput. The actual weight used during network operation is determined by the degree of protection afforded to the primary link. We begin by formulating the problem for a channel with a general number of states. We then study a two-state model where we characterize a scheme that spans the boundary of the primary-secondary rate region. Moreover, we study a three-state model where we derive the optimal strategy using dynamic programming. We also extend our two-state model to a two-channel case, where the secondary user can decide to transmit on a particular channel or not to transmit at all. We provide numerical results for our optimal strategies and compare them with simple greedy algorithms for a range of primary channel parameters. Finally, we investigate the case where some of the parameters are unknown and are learned using hidden Markov models (HMM).
\end{abstract}

\section{Introduction}
\label{ch:introduction}

Cognitive radio technology is a solution to the problem of spectrum under-utilization caused mainly by static spectrum allocation. In cognitive radio networks, the licensed users coexist with cognitive users, also known as the secondary or unlicensed users. Primary users are licensed users who are assigned to certain channels, and they have the right to transmit over the band of spectrum assigned by regulatory bodies in their respective countries. If other unlicensed, or secondary, users want to share the spectrum, then they must use the spectrum when it is unused by the primary users and/or when they can limit the interference they cause on the primary receivers below a certain specified level. The secondary users attempt to utilize the resources unused by the primary users adopting procedures that aim at protecting the primary network from service interruption and interference.

There has been interest in schemes that make use of the feedback of the primary link to predict the behavior of the primary user in the future and, in the case of primary channel temporal correlation, to gain knowledge about the channel between primary transmitter and receiver (e.g., \cite{Bits through ARQs}, \cite{Cognitive Power Control} and \cite{Exploiting Hidden Power-Feedback}). In \cite{Bits through ARQs}, the secondary user observes the automatic repeat request (ARQ) feedback from the primary receiver. The primary user achieved packet rate can be calculated by counting the ACK feedback messages. The cognitive radio's objective is to maximize secondary throughput under the constraint of guaranteeing a certain packet rate for primary user. The main difference between our work and \cite{Bits through ARQs} is that in \cite{Bits through ARQs} there is no use of the possible channel correlation across time, whereas we assume that the primary channel state follows a Markov chain. The cognitive transmitter can hence exploit the ACK/NACK feedback messages to predict the primary channel state during the next transmission phase. In \cite{Cognitive Power Control}, assuming a temporally correlated channel between the primary transmitter and receiver, the cognitive transmit power is adjusted based on primary channel state information (CSI) feedback. A real-time fading channel model is assumed rather than a  channel with a finite number of states as we consider and discuss below. However, the computation of the optimal procedure in \cite{Cognitive Power Control} is computationally prohibitive. In \cite{Exploiting Hidden Power-Feedback}, the cognitive radio adopts an active sensing protocol where, prior to the sensing, the cognitive radio generates a temporary jamming signal to deliberately interfere with the primary user message. If the primary channel is indeed active, the primary network will increase its transmit power upon receiving the jamming signal. The cognitive radio will receive the power-boosted primary user signal that is more easily detectable. Although this hidden power-feedback loop should increase the reliability to detect the presence of the primary user, in our proposed scheme we assume a passive cognitive radio and the primary link is always active where the cognitive radio can also transmit during the erasure states as we will show below.

There has been a series of recent work on cognitive MAC for opportunistic spectrum, e.g., \cite{Zhao}, \cite{Opportunistic Spectrum Access}, \cite{Ding}, \cite{marco1}, \cite{marco2}, and \cite{marco3}. In \cite{Zhao}, the primary user activity remains fixed over the duration of a slot and switches between idle and active states according to a two-state Markovian process. The channel between the primary transmitter and receiver is not considered, and the feedback used to predict the channel availability is provided by the secondary receiver. However, in our work, the primary channel is considered and its state changes between different time slots and the feedback is provided by the primary receiver not the secondary one. In \cite{Opportunistic Spectrum Access}, the work in \cite{Zhao} is extended to account for energy consumption and spectrum sensing duration optimization. In \cite{Ding}, the authors focus on the ARQ messages used in primary user data-link-control which are overheard by the secondary transmitter, which can optimize its access policy by assessing primary user reception quality. The primary channel is assumed to be of fixed quality resulting in two fixed and known packet error rates corresponding to the presence and absence of secondary user transmissions. The difference between our work and that in \cite{Ding} is that in our work, the primary channel has different channel qualities depending on the primary channel state. The more the channel states the primary channel have, the more the model will be more realistic and the more efficient utilization of the channel recourses. In \cite{marco1}, the coexistence of two unlicensed links is considered, where one link interferes with the transmission of the other with quasi-static fading and in the absence of channel state information at the nodes. The problem is formulated as a Partially Observable Markov Decision Process (POMDP) and a greedy solution is proposed. In \cite{marco2}, the secondary source is allowed to superpose its transmissions over those of the primary source. The secondary source aims to maximize its own throughput, while guaranteeing a bounded performance loss for the primary source. \cite{marco3} goes along the same line, but considers secondary users cooperation. In this type of problems the framework of Partially Observable Markov Decision Processes (POMDP) is usually needed given the uncertainties of the quality of the primary link, and of the primary user activity as a result of sensing errors \cite{POMDP1,POMDP2,Dynamic Programming}.

If we assume that the transition probabilities of the underlying Markov chain of the primary channel is unknown, we use a Hidden Markov Model framework to learn these transition probabilities. The problem of HMM arises when the actual system state is unknown. HMM is widely used in many applications. The detailed procedure of estimating transition probabilities is provided in \cite{Hidden_Markov}. In \cite{Ghosh} and \cite{Ghosh2}, HMM has been used for spectrum sensing. The sub-band occupancy at any time instant can be considered as a state, which can be either free (unoccupied by a primary user) or busy (occupied by a primary user). The states of a sub-band are monitored over $L$ consecutive time periods. There are two possible cases discussed in \cite{Ghosh}. The first option is to deal with the situation when the parameters of the HMM and error probabilities are known. Viterbi algorithm is used to overcome the computational complexity in the likelihood solution. The second alternative is to deal with the situation when the parameters of Markov chain and the error probabilities are unknown. Expectation-Maximization (EM) algorithm is used to estimate these parameters.

Note that HMM has been used in the context of cognitive radios like in \cite{Akba} and \cite{Quickest Spectrum Detection}. In \cite{Akba}, HMM is used to model and predict the spectrum occupancy of licensed radio bands.  In \cite{Quickest Spectrum Detection}, HMM is used for quickest spectrum detection for cognitive radio where a frequency sweeping device sweeps the wide-band spectrum and the samples of the wide-band power spectrum density (PSD) are fed into different HMMs sequentially. In our work, the ARQ  is considered as the secondary user observation vector, which can be considered as an input to HMM to estimate the transition probabilities. The quality of estimation depends on the length of the observation vector.

In this paper, we assume that the primary user is working in a saturation regime where it always has data to transmit over the primary link. It sends one packet in each time slot, and receives an ACK or NACK feedback from its receiver at the end of the time slot. The feedback is received correctly by both the primary and secondary transmitters. The channel between the primary transmitter and receiver is modeled as a Markov process with a finite number of states where the channel quality and hence the probability of correct reception is determined by the state. The state of the channel does not change over a slot. The channel may switch states at the beginning of each slot according to the transition probabilities of the Markov process. We laos study the special cases when the primary link follows a two state and a three state Markov process. The cognitive user exploits the ACK/NACK feedback from the primary receiver to predict the quality of the primary link. Note that, similar to other papers, this is also a POMDP problem and we can use techniques similar to other approaches. We assume that one secondary user operates on one primary channel. This is a widely used assumption in the literature. The system may have a number of orthogonal primary channels where each can be potentially used by one secondary terminal. At the beginning of each time slot, the secondary user decides whether to remain silent and listen to primary user feedback, or to carry out transmission. The objective is to maximize the weighted sum throughput of both the primary and secondary links.

This optimization problem has an exploration-exploitation tradeoff. The tradeoff is between the choice of the cognitive user activity which maximizes the secondary user throughput, and that which gives the secondary user knowledge about the channel state information of the primary link through the ARQ feedback. The correlation between the channel states, via the Markov process, is what enables us to have this trade-off. A similar trade-off can be achieved if there is correlation between primary user activity in consecutive time slots \cite{marco2}. A more general model would be to assume both correlations. Note that the maximization of the weighted sum throughput can be a proxy for the optimization problem of maximizing the secondary throughput under a constraint on the primary throughput \cite{dual problem}. Note however that solving the constrained optimization problem is more general.

Our contributions in this paper are as follows. We solve the weighted sum throughput maximization problem via modeling it as a dynamic programming problem, and employ Bellman's equation \cite{Dynamic Programming} to arrive at the optimal strategy. For the two state case, we prove that the optimal policy is a threshold based policy in the belief of the state of the channel. This threshold can be obtained numerically. Moreover, when the discounting factor of the dynamic programming problem approaches unity we obtain a closed form solution to the weighted sum throughput optimization problem, and can find exact description of the strategy that maximizes this throughput for any weight. Note that changing the weight spans the boundary of the primary-secondary rate region.
We then extend the two state single channel case to the case of a channel with three states and to the multiple channel case. We obtain numerical solutions to these problems. We then, and rather than assuming knowledge of the transition probabilities of the above model, investigate the case  where some of the primary network state transition probabilities are unknown, and we show how to estimate these parameters using Hidden Markov Model (HMM) formulation.

One of the advantages of our scheme is that the ARQ feedback can capture the temporal correlation in the channel. The cognitive user can access the primary channel in two cases, when the primary channel quality is relatively high (primary user can transmit successfully regardless of cognitive user activity) and when its quality is very low (primary user transmission fails whether secondary user is active or not). This advantage cannot be captured in schemes employing spectrum sensing only.

The paper is organized as follows. The general multi-state system model and assumptions are described in Section \ref{sec:Two-state System model}. We also introduce in this section the two-state, Gilbert-Elliot, three-state system models as special cases and the secondary user optimal strategy is obtained using dynamic programming. In section \ref{sec:Multichannel}, we extend our formulation to the multi-channel case and specifically address the two channel case. When the secondary user have the chance to choose a channel from multiple primary channels to transmit on, the secondary user should choose the channel that maximizes the reward function as we illustrate in Section \ref{sec:Multichannel}. The framework of Hidden Markov Model (HMM) is used in Section \ref{sec:Hidden Markov model} to estimate the primary network transition probabilities if it is unknown given the feedback observations in the form of ARQ feedback. Numerical results are presented in Section \ref{sec:Simulation Results}. Our work is concluded in Section \ref{sec:Conclusion}.

\section{System Model and Proposed Approach}

\label{sec:Two-state System model}
\begin{figure}[!ht]
\centering
  \includegraphics[width=0.5\columnwidth]{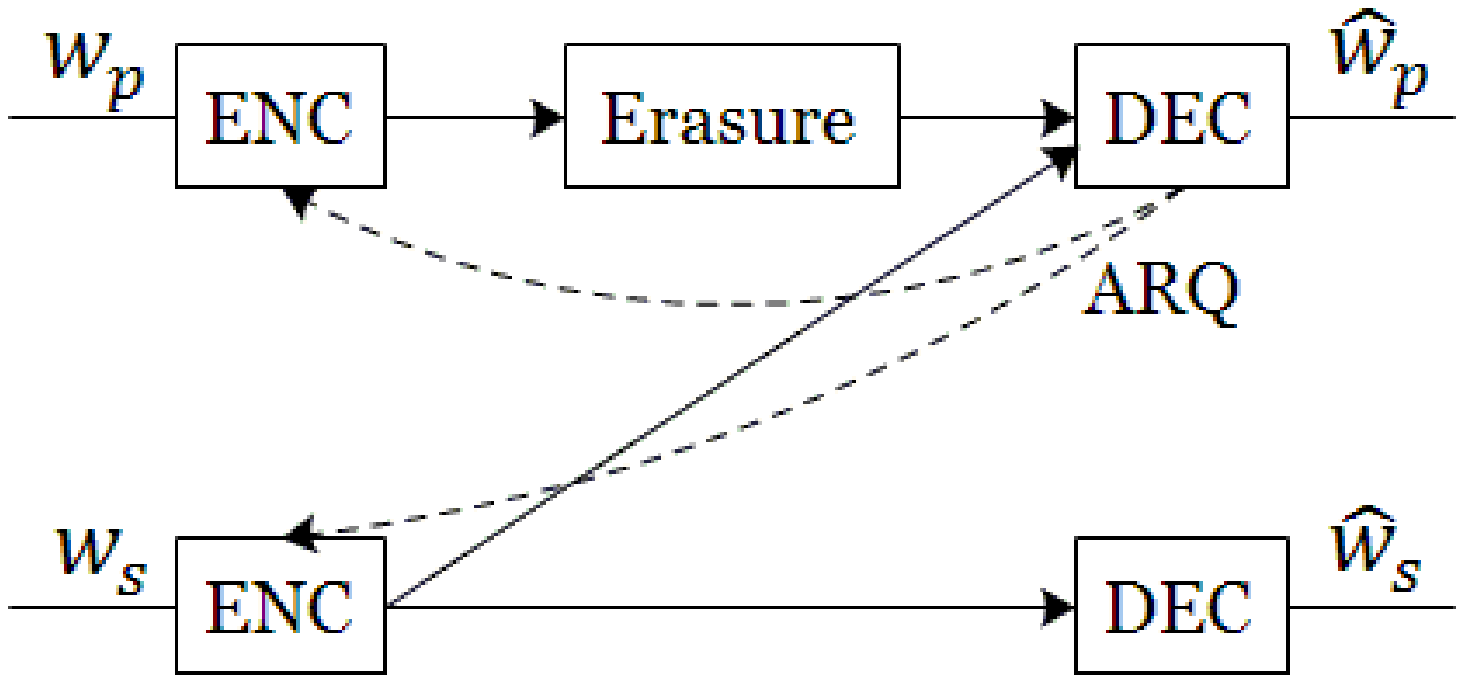}\\
  \caption{Z-interference erasure channel model. The secondary transmitter, when active, causes interference on the primary receiver. The secondary receiver, on the other hand, is shadowed from the primary transmitter, thereby suffering no interference from it. ENC is the channel encoder, whereas DEC is the receive decoder.}\label{fig:system_model2}
\end{figure}

Our proposed model assumes that we have one primary link and one secondary link. An illustration of the setup is provided in Figure \ref{fig:system_model2}. We are concerned with the Z-interference channel model \cite{Z channel} where the interference from the primary transmitter on the secondary receiver is ignored. The Z-interference channel models important applications such as the interference caused by macro-cell users on femto-cell receivers, which is known in the literature as the ``loud neighbor'' problem. In our context, the primary terminals may be close to one another and use small transmission power, whereas the cognitive terminals may be far from one another and use high power for communication causing considerable interference on the primary link.

We assume that the primary link is always active, i.e., primary transmitter is working in saturation regime. The primary link follows a Markov model with $S$ states; $\{1,2,...,S\}$. Each state of the $S$ states describes a certain quality for the primary channel. The secondary user is aware of the state transition matrix $\prod$ of this Markov chain in addition to the probabilities of success and failure for primary transmission in each of these states. Table II shows these probabilities for state $i$, $i=\{1,2,...,S\}$
\begin{table}[ht]
\label{eqn: table}
	\centering
		\begin{tabular}{l||l|l|}
		    & Sec. silent & Sec. transmit\\
			\hline\hline
			Prob. of success & $a_i$ & $b_i$\\
			\hline	
			Prob. of failure & $1-a_i$ & $1-b_i$\\
			\hline
		\end{tabular}
	\caption{Probability of successful delivery of primary user packets}
\end{table}

Our objective is to choose the transmission strategy that maximizes the expected value of the average weighted sum throughput $R$ given by
\begin{equation}
\label{eqn: 1st eqn}
R=\lim_{K \to \infty}\frac{1}{K}\sum_{n=1}^{K}\Bigl(wR_{\rm p}(n)+\left(1-w\right)R_{\rm s}(n)\Bigl)
\end{equation}
where $R_{\rm p}(n)$ and $R_{\rm s}(n)$ are the primary and secondary throughput in time slot $n$ respectively, and $0\leq w \leq 1$ is a weighting factor that determines the relative importance of the two rates. In order to protect the primary user from interference and service interruption, parameter $w$ can be chosen close to one. Note that the maximization of the weighted sum throughput can be a proxy for the optimization problem of maximizing the secondary throughput under a constraint on the primary throughput \cite{dual problem}. Note however that solving the constrained optimization problem is more general than solving (1) \cite{Boyd}. Consider for example a convex primary-secondary throughput region, and consider the primary throughput to be constrained to a certain point on the boundary of this region. The maximum secondary throughput that can be obtained is the one that solves the optimization of the weighted sum throughput where the ``weight" produces a line tangent to the boundary of the region at this certain point. The weighted sum formulation hence resembles imposing a constraint on the primary throughput. Now this is tougher and more meaningful that the typical interference constraint. An interference constraint is concerned with the total interference power at the primary receiver regardless of the quality of the primary channel. On the other hand, a throughput constraint is related to both the interference imposed on the primary receiver and the quality of the primary transmitter-receiver link. If the primary link is in a good state, the primary receiver may tolerate more interference from the secondary user without an appreciable degradation in performance.

Let $r_{\rm p}$ be the primary user reward, if the primary user succeeds to transmit one packet through the binary erasure channel. Let $r_{\rm s}$ be the secondary user reward, if the secondary user decides to transmit a packet. Note that we can take account of any possible packet loss in the secondary channel when we calculate the value of $r_{\rm s}$.

The belief is the secondary user evaluation of the probability of the actual primary link state in the next time slot given the history of actions and observations. The secondary belief is given by the vector $\overrightarrow{p}=(p_1 p_2 ... p_S)$ with $\sum_{k=1}^{S} p_k=1$, where $p_k$ is the probability that the primary channel is in state $k$ at the beginning of a time slot given all the previous secondary user's actions and observations. The belief is updated by the secondary user in each time slot on the basis of its chosen action during the time slot, the primary ARQ feedback that it observes, and the primary channel state transition probabilities. Specifically, the secondary user updates this belief of the primary state according to Markovian update equation $\overrightarrow{p}_n=\overrightarrow{p}_{{n-1}{(updated)}}\prod$, where $\overrightarrow{p}_n$ is the belief vector at time slot $n$, and $\overrightarrow{p}_{{n-1}{(updated)}}$ is calculated using the ACK or NACK overheard from the primary transmitter according to one of the following four
cases depending on the action taken by the secondary user and the corresponding outcome.

Case 1: Secondary user transmits and a NACK is received from the primary network. Therefore, the secondary belief of the primary channel state during the next time slot is
\begin{eqnarray}
  p_{i(updated)} (1)= \frac{P(NACK|i)p_i}{\sum _i P(NACK|i)p_i} = \frac{(1-a_i)p_i}{\sum _i (1-a_i)p_i}
\end{eqnarray}
where $p_{i(updated)}(l)$ is the $i_{th}$ element of $\overrightarrow{p}_{{n-1}{(updated)}}$ for the $l_{th}$ case and $P(NACK | i)$ is the probability of receiving a NACK given the primary channel state is $i$.

Case 2: Secondary user transmits and an ACK is received from the primary network. Therefore, the secondary belief of the primary channel state during the next time slot is
\begin{eqnarray}
   p_{i(updated)}(2)  = \frac{P(ACK|i)p_i}{\sum _i P(ACK|i)p_i} = \frac{a_ip_i}{\sum _i a_ip_i}
\end{eqnarray}

Case 3: Secondary user is idle and a NACK is received from the primary network. Therefore, the secondary belief of the primary channel state during the next time slot is
\begin{eqnarray}
  p_{i(updated)}(3) = \frac{(1-b_i)p_i}{\sum _i (1-b_i)p_i}
\end{eqnarray}

Case 4: Secondary user is idle and an ACK is received from the primary network. Therefore, the secondary belief of the primary channel state during the next time slot is
\begin{eqnarray}
   p_{i(updated)}(4)  = \frac{b_ip_i}{\sum _i b_ip_i}
\end{eqnarray}

Then, depending on the case, we have $\overrightarrow{p}_n(l)=\overrightarrow{p}_{{n-1}{(updated)}}(l)\prod$

If the secondary user decided to remain silent, the expected immediate gain in the next time slot can be calculated as:
\begin{equation}
G_1(\overrightarrow{p}) = wr_{\rm p}P(ACK|idle) = wr_{\rm p}(\sum _i a_ip_i)
\end{equation}
where $P(ACK|idle)$ is the probability of receiving an ACK given the secondary user is idle.
If the secondary user is transmitting, the expected immediate gain in the next time slot can be calculated as:
\begin{equation}
G_2(\overrightarrow{p}) = (1-w)r_{\rm s} + wr_{\rm p}P(ACK|transmit) =  (1-w)r_{\rm s} + wr_{\rm p}(\sum _i b_ip_i)
\end{equation}

Thus, the expected immediate reward is:
\vspace{0.15cm}
\\
$R(k,\overrightarrow{p},a) = \left\{ \begin{array}{rcl}
{G_1(\overrightarrow{p})} & \mbox{if the secondary user is silent in time slot $k$}
 \\ {G_2(\overrightarrow{p})} & \mbox{if the secondary user is active in time slot $k$}
\end{array}\right.$
\vspace{0.15cm}
\\
The optimal strategy is the strategy that maximizes the following discounted reward function
\begin{equation}
E\left\{{\sum^{K+k-1}_{n=k} \alpha^{n-k}*R(n,\overrightarrow{p}_n,a_n) \hspace{2mm}|\hspace{2mm} \overrightarrow{p}_k=p}\right\}
\end{equation}
$V^K(\overrightarrow{p})$ satisfies the following Bellman equation \cite{Dynamic Programming}:
\begin{equation}
\label{eqn: itrative 3}
\begin{split}
V^K(\overrightarrow{p})= {\rm max} \bigg\{ &wr_{\rm p}P(ACK|idle)+\alpha P(ACK|idle)V^{K-1}(\overrightarrow{p}(4)) \\
  & + \alpha (1-P(ACK|idle))V^{K-1}(\overrightarrow{p}(3)),  \\
  &  (1-w)r_{\rm s}+wr_{\rm p}P(ACK|transmit)+\alpha P(ACK|transmit)V^{K-1}(\overrightarrow{p}(2)) \\
   & + \alpha (1-P(ACK|transmit))V^{K-1}(\overrightarrow{p}(1)) \bigg\}
\end{split}
\end{equation}
where
\begin{equation}
V^1(\overrightarrow{p})= {\rm max} \{wr_{\rm p}P(ACK|idle), (1-w)r_{\rm s}+wr_{\rm p}P(ACK|transmit)\}
\end{equation}
When $K=\infty$, $V(p)$ denote the maximum achievable discounted reward function.
$V(p)$ satisfies the following Bellman equation \cite{Dynamic Programming}:
\begin{equation}
\begin{split}
V(p)= {\rm max} \bigg\{ &wr_{\rm p}P(ACK|idle)+\alpha P(ACK|idle)V(\overrightarrow{p}(4))+\alpha (1-P(ACK|idle))V(\overrightarrow{p}(3)), \\
  &  (1-w)r_{\rm s}+wr_{\rm p}P(ACK|transmit)+\alpha P(ACK|transmit)V(\overrightarrow{p}(2)) \\
   & + \alpha (1-P(ACK|transmit))V(\overrightarrow{p}(1)) \bigg\}
\end{split}
\end{equation}

The solution can be obtained via Dynamic Programming. We defer the explanation of our algorithm to
the following subsections where we introduce some special cases to better explain our approach.

\subsection{Erasure/Non-Erasure Channel Model}

\begin{figure}[!ht]
	\centering
  \includegraphics[width=0.5\columnwidth]{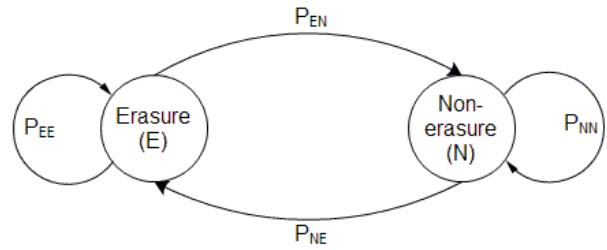}\\
  \caption{Two-state Markov model.}\label{fig:states2}
\end{figure}

The primary link follows a two-state Markov chain. This is a special case of the model described above, where $S$=2, and the $a_i$'s and $b_i$'s are all '1' or '0'. In particular, we assume that  during each time slot, the primary link is either in an erasure state $E$ where primary user transmission always fails or a non-erasure state $N$ where primary user transmission is successful when there is no interference. It switches states from one time slot to the next according to a Markovian process as shown in Figure \ref{fig:states2}. The process is specified by two parameters $P_{\rm EE}$ and $P_{\rm NE}$, where $P_{\rm EE}$ is the probability that the primary network is in erasure in the next time slot given that it is in erasure state in the current slot, and $P_{\rm NE}$ is the probability that the primary network is in erasure in the next time slot given that it is in non-erasure state in the current slot.  We assume that the transition probabilities of the Markov chain are known a priori but we introduce a technique for estimating them in Section \ref{sec:Hidden Markov model}. The transition matrix $P$ which includes the transition probabilities is given by
\[
P =
\left[ {\begin{array}{cc}
 P_{\rm EE} & P_{\rm EN}  \\
 P_{\rm NE} & P_{\rm NN}  \\
 \end{array} } \right]
\]
Hence, the stationary probabilities of being in erasure and non-erasure for primary network are $P(E)=\frac{P_{\rm NE}}{P_{\rm NE}+P_{\rm EN}}$ and $P(N)=\frac{P_{\rm EN}}{P_{\rm NE}+P_{\rm EN}}$, respectively. Note that, the erasure state causes the primary transmission to fail, while the non-erasure state results in successful packet delivery to the primary receiver \emph{only} when there is no interference from the secondary transmitter. That is, if the cognitive user decides to transmit in the non-erasure state, its transmission causes the erasure of the primary user packet.

As long as the cognitive radio is idle, it can eavesdrop on the primary user ARQ through which the secondary transmitter can detect the current state of the primary link and, consequently, decide the erasure probability of the next state whether $P_{\rm EE}$ or $P_{\rm NE}$ based on the ARQ feedback. However, if the cognitive radio decides to transmit, it causes the primary user packet to be erased. The secondary user then overhears a negative acknowledgment (NACK) from the primary receiver no matter what the state of the primary channel is. Thus when the cognitive user transmits, it becomes unaware of the primary link state because of the collision it causes with the primary transmitted data. The primary receiver will hence not be able to decode the data, and will send a negative acknowledgment to its transmitter irrespective of its actual channel state. Hence, the ACK/NACK feedback provides no information because the secondary transmitter is unable to decide whether the negative ARQ feedback is actually due to the collision or to the fact that the primary state was in erasure.

The belief here is defined as the probability that primary link is in erasure state at the very beginning of the time slot from the secondary terminals's perspective given the history of its actions and observations. Hence, the expected primary throughput in time slot $k$ as estimated by the secondary transmitter can be given by $r_{\rm p}\left(1-p_k\right)$. The belief is updated from one time slot to another using the Markov property according to the following.
\vspace{0.15cm}
\\
$p_k = \left\{ \begin{array}{cc}
{P_{\rm EE}} & \mbox{if current state is erasure}
 \\{P_{\rm NE}} & \mbox{if current state is non-erasure}  \\
{p_{k-1}P_{\rm EE}+(1-p_{k-1})P_{\rm NE}} & \mbox{if current state is unknown}
\end{array}\right.$
\vspace{0.15cm}
\\
Note that the third possibility occurs when the secondary user transmits. The expected secondary throughput in time slot $k$ is as before given by $r_{\rm s}$

The optimal strategy and maximum throughput can be derived using one of the following approaches.
\subsubsection{Dynamic Programming}

We propose to use dynamic programming techniques to arrive at the optimal strategy. The secondary user opportunistically accesses the channel by first synchronizing with the slot structure of primary network. The goal of the secondary user is to transmit during the erasure states, and allow the primary user to transmit during the non-erasure states in order to maximize the sum throughput. The main challenge is that the secondary user cannot know the next state exactly. It has to operate on the basis of its belief $p_k$.  At the beginning of each time slot, and based on previous actions and observations, the secondary user can calculate the probability $p_k$. Dynamic programming can then be used to find the optimal strategy.  The decision at each value of $p_k$ maximizes the expected reward function and hence the optimal strategy at time slot $k$ is the strategy that maximizes the following discounted reward function \cite{Opportunistic Spectrum Access}
\begin{equation}
\label{eqn: 7th eqn}
V^{K}\left(p\right)=\max_{a_k,a_{k+1},...a_{K+k+1}}E\Bigg\{\sum^{K+k-1}_{n=k} \alpha^{n-k}R\left(n,p_n,a_n\right) \hspace{2mm}|\hspace{2mm} p_k=p\Bigg\}
\end{equation}
\noindent where $0 \leq \alpha < 1$ is a discounting factor, $K \in \{1,2,..\}$ is the control horizon, and $a_n$ is the action taken at time $n$. As $\alpha$ decreases, the secondary user puts more emphasis on its short-term future gains. The term $R\left(n,p_n,a_n\right)$ is the reward accrued at instant $n$ when the belief is $p_n$ and the action taken is $a_n$. In this paper, the reward is given by
\begin{equation}
R\left(n,p_n,a_n\right)=wR_{\rm p}\left(n,p_n,a_n\right)+\left(1-w\right)R_{\rm S}\left(n,p_n,a_n\right)
\end{equation}
\noindent where $R_{\rm p}$ and $R_{\rm s}$ are the primary and secondary throughputs, respectively.

Note that at a certain time instant and a certain $p_k$ value, the path through the "decision tree" depends on the decision taken-on whether to transmit or not. For that particular $p_k$ value at any other time instant, the optimal decision would not change. Hence we eliminate the subscript $k$ because the expected future reward depends only on the value of $p$ regardless of the index of the time slot.

If we assume an infinite horizon optimization, and through a dynamic programming argument, the state of the system can be fully parameterized by the belief that the channel is in erasure the next time slot, $p$, where we dropped the time dependence. Hence the action taken by the cognitive user depends only on $p$. The belief state $p$ can be updated according to one of the following three cases depending on the action taken by the secondary user and the corresponding outcome. We follow here the notation presented in \cite{Opportunistic Spectrum Access}.

Case 1: Secondary user is silent and a positive acknowledgment (ACK) is received from the primary network. The ACK implies that primary network has been in the non-erasure state, and primary receiver has succeeded in decoding the packet. Therefore, the secondary belief that the channel would be in erasure during the next time slot is
\begin{equation}
p(1)=P_{\rm NE}
\end{equation}
where $p(1)$ is the update expression for $p$ for the $l$th case. Each case is a certain combination of secondary user decision and observation.

Case 2: Secondary user is silent and a negative acknowledgment (NACK) is received from the primary network. This implies that primary network has been in erasure state and the sent packet has not been delivered successfully. Thus,
\begin{equation}
p(2)=P_{\rm EE}
\end{equation}

Case 3: Secondary user transmits. The probability of erasure is updated by the Markovian property as follows
\begin{equation}
p(3)=pP_{\rm EE}+\left(1-p\right)P_{\rm NE}
\end{equation}
\noindent We denote the weighted expected instantaneous throughput when the secondary user listens by $G_1\left(p\right)$ which is given by
\begin{equation}
G_1\left(p\right)=wr_{\rm p}\left(1-p\right)
\end{equation}
\noindent For the case of the secondary user is transmitting, we denote the weighted expected throughput by
\begin{equation}
G_2\left(p\right)=\left(1-w\right)r_{\rm s}
\end{equation}

A greedy scheme would just compare $G_1(p)$ with $G_2(p)$ and if $G_2(p)>G_1(p)$, the secondary user decides to transmit, otherwise it remains silent. The expected instantaneous reward is:
\begin{equation*}
R(k,p,a) = \left\{ \begin{array}{rcl}
{G_1(p)} & \mbox{if the secondary user is silent at time slot $k$}
 \\ {G_2(p)} & \mbox{if the secondary user is active at time slot $k$}
\end{array}\right.
\end{equation*}

Following the definitions in \cite{Opportunistic Spectrum Access}, let $V^K\left(p\right)$ denote the maximum achievable discounted reward function obtained by maximizing Equation (\ref{eqn: 7th eqn}). When $K<\infty$, $V^K\left(p\right)$ satisfies the following Bellman equation \cite{Dynamic Programming}:
\begin{equation}
\label{eqn: itrative bellman}
\begin{split}
V^K(p)= {\rm max} \bigg\{ &wr_{\rm p}(1-p)+\alpha (1-p)V^{K-1}\left(p(1)\right)+\alpha pV^{K-1}\left(p(2)\right),\\
   & (1-w)r_{\rm s}+\alpha V^{K-1}(p(3)) \bigg\}
\end{split}
\end{equation}
where $wr_{\rm p}(1-p)$ and $\alpha (1-p)V^{K-1}\left(p(1)\right)+\alpha pV^{K-1}\left(p(2)\right)$ correspond to the expected immediate primary user reward and total discounted future reward respectively if the secondary user does not transmit, and $(1-w)r_{\rm s}$ and $\alpha V^{K-1}(p(3))$ correspond to the expected immediate secondary
reward and total discounted future reward respectively if the secondary user transmits.
\begin{equation}
V^1(p)= {\rm max} \left\{wr_{\rm p}(1-p), (1-w)r_{\rm s}\right\}
\end{equation}
When $K=\infty$, $V^K(p)=V^{K-1}(p)=V(p)$ which satisfies the following Bellman equation:
\begin{equation}
\label{eqn: steady state bellman}
V(p)= {\rm max} \bigg\{ wr_{\rm p}(1-p)+\alpha (1-p)V\left(p(1)\right)+\alpha pV\left(p(2)\right), (1-w)r_{\rm s}+\alpha V(p(3)) \bigg\}
\end{equation}
$V(p)$ can be obtained iteratively. The optimal strategy is obviously a function of $p$. We show in Appendix \ref{app:Properties of the Function} that the optimal strategy is a threshold based policy, i.e. the secondary user will transmit if $p>p_{th}$. In the next subsection, we present a closed form expression for the maximum throughput and optimal strategy when the discounting factor $\alpha$ tends to unity.
\subsubsection{Closed Form Solution when $\alpha \rightarrow 1$}

We assume that $P_{\rm EE}>P_{\rm NE}$ making the belief $p_k$ a monotonic function with time as long as
 the secondary user is transmitting \cite{Restless Bandit Problems}. This can be readily seen by solving the first order difference equation governing the evolution of $p_k$ to obtain
\begin{equation}
 p_k=(P_{\rm EE}-P_{\rm NE})^{K}p_{k-K}+\left[1-(P_{\rm EE}-P_{\rm NE})^{K}\right]P\left(E\right)
\end{equation}
where $P_{\rm EE}$ and $P_{\rm NE}$ are Markov chain transition probabilities, $P\left(E\right)$ is the steady state probability of being in an erasure state $E$ where $P\left(E\right)=\frac{P_{\rm NE}}{P_{\rm NE}+P_{\rm EN}}$, and $p_k$ is the probability of being in erasure in time slot $k$.\\
Using this equation, we can find the probability of erasure in time slot $k$, $p_k$ as a function of probability of erasure in time slot $(k-K)$, $p_{k-K}$.\\
For example, if we consider a single time slot shift $(K=1)$, we will get the basic Markovian property.
\begin{equation}
 p_k=p_{k-1}P_{\rm EE}+\left[1-p_{k-1}\right]P_{\rm NE} \nonumber
\end{equation}
It is clear that if $P_{\rm EE}-P_{\rm NE}>0$, the belief $p_k$ is a monotonic function with time, otherwise the term $(P_{\rm EE}-P_{\rm NE})^{K}$ oscillates between positive and negative values.

If the inequality $wr_{\rm p}\left(1-P_{\rm EE}\right)>(1-w)r_{\rm s}$ is satisfied, this implies that the inequality $wr_{\rm p}\left(1-P_{\rm NE}\right)>(1-w)r_{\rm s}$ is also satisfied, since $P_{\rm EE}$ is greater than $P_{\rm NE}$. Thus, from the two previous inequalities, we can deduce that the optimal secondary user strategy is to listen always because the expected primary throughput is greater than the expected secondary throughput regardless of the actual system state, whether it is $E$ or $N$. Similarly, if $wr_{\rm p}\left(1-P_{\rm NE}\right)<(1-w)r_{\rm s}$, the optimal secondary strategy is to transmit always. For any other condition, the optimal secondary strategy is as follows. The secondary transmitter listens as long as an ACK is received because $wr_{\rm p}\left(1-P_{\rm NE}\right)>(1-w)r_{\rm s}$ and in that case maximizing the throughput in the next time slot is optimal since we do not affect future decisions as the secondary user will make use of the knowledge of the next state while it is silent. Once a NACK is received, the secondary user transmits $M$ consecutive packets. Thus, the maximization problem in (\ref{eqn: 7th eqn}) can be reduced to an optimization problem over the number of secondary user consecutive transmitted packets $M$.

\begin{figure}[!ht]
\centering
  \includegraphics[width=0.5\columnwidth]{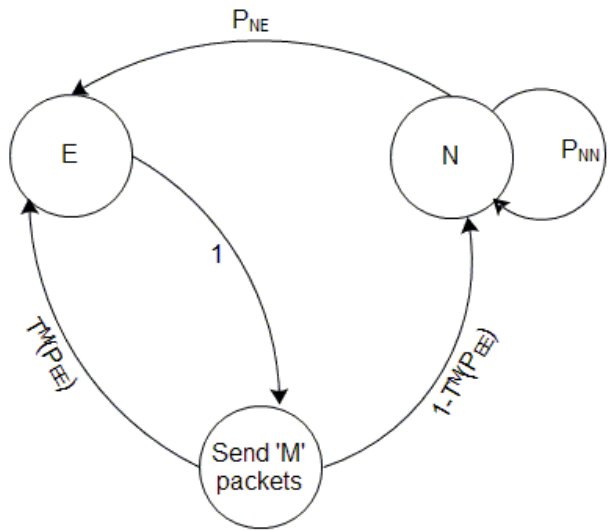}\\
  \caption{Closed Form Solution when $\alpha \rightarrow 1$.}\label{fig:3states1}
\end{figure}

It is hence possible to model the system as seen by the secondary user as a 3-state Markovian chain, where the secondary user either knows that the primary link is in erasure or non-erasure (while silent) or sends $M$-packets without actually knowing the exact state of the primary link. Figure \ref{fig:3states1} shows such a representation of our algorithm. The first state N represents the state that the secondary user is silent and that the primary channel state is in the non-erasure state. The secondary user overhears an ACK at the end of the primary transmission, and hence the probability to remain in the same state is $P_{\rm NN}$ and the probability to move to state E is $P_{\rm NE}$. The state E refers to the state where the  primary link in the erasure state. The secondary user receives a NACK at the end of the primary user transmission, and hence moves to the ``Send $M$ packets" state. The secondary user transmits $M$ consecutive packets, then the secondary user returns back to idle state and starts to listen to the feedback again and returns to state E or N based on this feedback.

When the Markov chain is in state $N$, the primary network achieves a throughput of $r_{\rm p}$. When it is in state $S$, the secondary user achieves a throughput of $Mr_{\rm s}$ as the system remains in this state for $M$ time slots.

In order to find an expression of the throughput as a function of $M$, we find the stationary distribution of each state of the Markov chain. Let the steady state probability of $N$, $E$ and $S$ states be $P^{ss}_{\rm N}$, $P^{ss}_{\rm E}$ and $P^{ss}_{\rm S}$ respectively, then
\begin{equation}
P^{ss}_{\rm N}=\frac{1-T^M (P_{\rm EE})}{1+2P_{\rm NE}-T^M (P_{\rm EE})}
\end{equation}
\begin{equation}
P^{ss}_{\rm E}=P^{ss}_{\rm S}=\frac{P_{\rm NE}}{1+2P_{\rm NE}-T^M (P_{\rm EE})}
\end{equation}
\noindent where $T^M (P_{\rm EE})$ is the probability of erasure in time slot $k$ given that the state of Markov chain in time slot $k-M$ was erasure. We can find $T^M (P_{\rm EE})$ from the two-state Markov chain:
\begin{equation}
T^M (P_{\rm EE})=\frac{P_{\rm NE}+(P_{\rm EE}-P_{\rm NE})^{(M+1)}\, (1-P_{\rm EE})}{1+P_{\rm NE}-P_{\rm EE}}
\end{equation}
\noindent Recall that we assume a positively correlated channel with $P_{\rm EE}>P_{\rm NE}$.

The average primary throughput $R_{\rm p}$ and the average secondary throughput $R_{\rm s}$ can be written as:
\begin{equation}
R_{\rm p}=\frac{r_{\rm p}P^{ss}_{\rm N}}{P^{ss}_{\rm N}+P^{ss}_{\rm E}+MP^{ss}_{\rm S}}
\end{equation}
\begin{equation}
R_{\rm s}=\frac{r_{\rm s}MP^{ss}_{\rm S}}{P^{ss}_{\rm N}+P^{ss}_{\rm E}+MP^{ss}_{\rm S}}
\end{equation}
\begin{equation}
\label{eqn: R(M)}
R(M)=wR_{\rm p}+(1-w)R_{\rm s}
\end{equation}
The problem now becomes  to find $M$ that maximize the average weighted sum throughput $R(M)$. We notice that the optimal value of $M$ depends on the weight $w$.
The throuhgput obtained using this scheme spans a number of points on the outer bound of the rate region with the optimal values of $M$ ($M$ is integer) that maximize the weighted sum throughput. The outer bound of the rate region here is piecewise linear, and can be achieved by time division multiplexing between the different integer values of $M$.

Two remarks are in order here:
\\1. We show in Appendix \ref{app:Properties of the Function} that the optimal strategy is a threshold-based policy on the belief $p_k$, and since the belief $p_k$ is monotonic with $M$, finding the threshold amounts to finding the value of $M$ that maximizes the throughput expression.
\\2. It can be shown that the throughput is a quasi-concave function of $M$, and through some algebraic manipulations, one can arrive at the value of $M$ that maximizes the throughput. This can be shown by subtracting $R(M)$ from $R(M+1)$. Treating $M$ as a continuous variable, we can show that this difference has only one positive finite root that is greater than or equal to unity. By finding this root, we find the value of the unique optimal $M$. This proof is shown in the Appendix \ref{app:Quasi-concavity}.

\subsection{Gilbert-Elliot Model}

\label{sec:limited feedback}
In the previous model, if the primary channel is in an erasure state, it will fail to deliver any packet with probability one, while in a non-erasure state it will succeed to deliver any packet with probability one. Here we use the more general model of a  good-bad Gilbert-Elliot model ($B$ and $G$ states), where the probability of successful delivery of a packet depends on the state of the channel but is not exactly 1 or 0. The probabilities also depend on whether the secondary user is transmitting or not, as discussed before. For example, when the primary channel is in the bad state $B$ and the secondary user is silent, there is a probability $\gamma_1$ of receiving an ACK (signifying correct reception of a packet) from the primary network. Table I gives the probabilities of such a channel model. This is the same as the general case when $S$=2, and here we set $\gamma_1=a_1$, $\gamma_2=b_1$, $\gamma_3=a_2$ and $\gamma_4=b_2$.
\begin{table}[ht]
\label{eqn: table}
	\centering
		\begin{tabular}{l||l|l|}
		    & Sec. silent & Sec. transmit\\
			\hline\hline
			Bad & $\gamma_1$ & $\gamma_2$\\
			\hline	
			Good & $\gamma_3$ & $\gamma_4$\\
			\hline
		\end{tabular}
	\caption{Probability of successful delivery of primary user packets}
\end{table}

Typically, the probabilities $\gamma_1$, $\gamma_2$ and $\gamma_4$ should be close to zero and the probability $\gamma_3$ should be close to one. The Erasure/Non-Erasure channel model corresponds to the case $\gamma_1=\gamma_2=\gamma_4=0$ and $\gamma_3=1$.

The state of the system can be fully parameterized by the belief that the channel is in $B$ state the next time slot, $p$. The action taken by the cognitive user depends only on $p$. The belief $p$ can be updated according to one of the following four cases depending on the action taken by the secondary user and the corresponding outcome.

Case 1: Secondary user transmits and a negative acknowledgment (NACK) is received from the primary network. Therefore, the secondary belief that the channel would be in $B$ state during the next time slot is
\begin{equation}
p(1)=P(B|NACK)P_{\rm BB}+(1-P(B|NACK))P_{\rm GB}
\end{equation}
where $P(B|NACK)$ is the probability of $B$ state in the current time slot given that the secondary user receives a NACK in the same time slot and $p(k)$ is the update expression for $p$ for the $k$th case. $P(B|NACK)$ can be obtained from Bayes' rule as follows.
\begin{eqnarray}
  P(B|NACK) & = & \frac{P(NACK|B)P(B)}{P(NACK|B)P(B)+P(NACK|G)P(G)} \nonumber \\
   & = & \frac{(1-\gamma_2)p}{(1-\gamma_2)p+(1-\gamma_4)(1-p)}
\end{eqnarray}

Case 2: Secondary user is transmitting and a positive acknowledgment (ACK) is received from the primary network. Thus,
\begin{equation}
p(2)=P(B|ACK)P_{\rm BB}+(1-P(B|ACK))P_{\rm GB}
\end{equation}
where
\begin{eqnarray}
  P(B|ACK) & = & \frac{P(ACK|B)P(B)}{P(ACK|B)P(B)+P(ACK|G)P(G)} \nonumber \\
   & = & \frac{\gamma_2p}{\gamma_2p+\gamma_4(1-p)}
\end{eqnarray}

Case 3: Secondary user is silent and a NACK is received from the primary network.
\begin{equation}
p(3)=P(B|NACK)P_{\rm BB}+(1-P(B|NACK))P_{\rm GB}
\end{equation}
where
\begin{eqnarray}
    P(B|NACK) & = & \frac{P(NACK|B)P(B)}{P(NACK|B)P(B)+P(NACK|G)P(G)} \nonumber \\
   & = & \frac{(1-\gamma_1)p}{(1-\gamma_1)p+(1-\gamma_3)(1-p)}
\end{eqnarray}

Case 4: Secondary user is idle and an ACK is received from the primary network.
\begin{equation}
p(4)=P(B|ACK)P_{\rm BB}+(1-P(B|ACK))P_{\rm GB}
\end{equation}
where
\begin{eqnarray}
  P(B|ACK) & = & \frac{P(ACK|B)P(B)}{P(ACK|B)P(B)+P(ACK|G)P(G)} \nonumber \\
   & = & \frac{\gamma_1p}{\gamma_1p+\gamma_3(1-p)}
\end{eqnarray}

The parameter $p$ characterizing the belief state is updated by one of the previous four conditions.
If secondary user is listening, the expected current gain can be calculated as:
\begin{equation}
G_1(p)=wr_{\rm p}P(ACK|idle)
\end{equation}
where $P(ACK|idle)$ is the probability of receiving an ACK given the secondary user is idle
\begin{eqnarray}
P(ACK|idle)=\gamma_1p+\gamma_3(1-p)
\end{eqnarray}
But if the secondary user is transmitting, the expected current gain is:
\begin{equation}
G_2(p)=(1-w)r_{\rm s}+wr_{\rm p}P(ACK|transmit)
\end{equation}
where $P(ACK|transmit)$ is the probability of receiving an ACK given the secondary user is transmitting
\begin{eqnarray}
P(ACK|transmit)=\gamma_2p+\gamma_4(1-p)
\end{eqnarray}
The expected current reward is:
\vspace{0.15cm}
\\
$R(k,p,a) = \left\{ \begin{array}{rcl}
{G_1(p)} & \mbox{if the secondary user is silent in time slot $k$}
 \\ {G_2(p)} & \mbox{if the secondary user is active in time slot $k$}
\end{array}\right.$
\vspace{0.15cm}
\\
The optimal strategy is the strategy that maximizes the following discounted reward function
\begin{equation}
E\left\{{\sum^{K+k-1}_{n=k} \alpha^{n-k}*R(n,p_n,a_n) \hspace{2mm}|\hspace{2mm} p_k=p}\right\}
\end{equation}
$V^K(p)$ satisfies the following Bellman equation \cite{Dynamic Programming}:
\begin{equation}
\label{eqn: itrative 3}
\begin{split}
V^K(p)= {\rm max} \bigg\{ &wr_{\rm p}P(ACK|idle)+\alpha P(ACK|idle)V^{K-1}(p(4)) \\
  & + \alpha (1-P(ACK|idle))V^{K-1}(p(3)),  \\
  &  (1-w)r_{\rm s}+wr_{\rm p}P(ACK|transmit)+\alpha P(ACK|transmit)V^{K-1}(p(2)) \\
   & + \alpha (1-P(ACK|transmit))V^{K-1}(p(1)) \bigg\}
\end{split}
\end{equation}
where
\begin{equation}
V^1(p)= {\rm max} \{wr_{\rm p}P(ACK|idle), (1-w)r_{\rm s}+wr_{\rm p}P(ACK|transmit)\}
\end{equation}
When $K=\infty$, $V(p)$ denote the maximum achievable discounted reward function.
$V(p)$ satisfies the following Bellman equation \cite{Dynamic Programming}:
\begin{equation}
\begin{split}
V(p)= {\rm max} \bigg\{ &wr_{\rm p}P(ACK|idle)+\alpha P(ACK|idle)V(p(4))+\alpha (1-P(ACK|idle))V(p(3)), \\
  &  (1-w)r_{\rm s}+wr_{\rm p}P(ACK|transmit)+\alpha P(ACK|transmit)V(p(2)) \\
   & + \alpha (1-P(ACK|transmit))V(p(1)) \bigg\}
\end{split}
\end{equation}
We show in Appendix \ref{app:limited feedback} that the optimum policy is a threshold based policy in the value of $p$. We solve Equation (\ref{eqn: itrative 3}) numerically. Specifically, we solve it iteratively via approximating the value function at a finite number of belief values on a grid. The value function is initialized and then (\ref{eqn: itrative 3}) is used to update it. For $p$ values not belonging to the grid, interpolation or extrapolation is used. After convergence, the secondary terminal decides whether to transmit or listen based on the term that maximizes $V(p)$ at each value of $p$.

\subsection{Three-state System Model}

\label{sec:Three-state system model}
In this subsection, we address the three-state model where the primary channel now follows a three-state Markov chain whose states are named Bad (B), Good (G) and Very good (Vg) with transition matrix $P$ where
\[
P =
\left[ {\begin{array}{ccc}
 P_{\rm BB} & P_{\rm BG} & P_{\rm BVg} \\
 P_{\rm GB} & P_{\rm GG} & P_{\rm GVg} \\
 P_{\rm VgB} & P_{\rm VgG} & P_{\rm VgVg} \\
 \end{array} } \right]
\]
If the secondary user is listening, the primary user can deliver its packet if the channel state is G or Vg. But if the secondary user is transmitting, the primary user transmission success is only in state Vg. This means that the primary and secondary users can both simultaneously transmit successfully in state Vg. Table III shows the primary packet success or failure during different channel states and secondary user activities.
\begin{table}[ht]
\label{eqn: table2}
	\centering
		\begin{tabular}{l||l|l|}
		    & Sec. silent & Sec. transmit\\
			\hline\hline
			B & fail & fail\\
			\hline	
			G & success & fail\\
			\hline
            Vg & success & success\\
			\hline
		\end{tabular}
	\caption{Primary packet delivery status during different secondary user activity}
\end{table}
\\We can also apply dynamic programming on that system with three channel states to arrive at the optimal decisions for the secondary user, whether to transmit or to listen, in any situation to maximize the weighted sum of the primary and secondary throughput.

Here we parameterize the belief state by two parameters $p$ and $q$, where $p$ is the probability that the primary network is in state G in the next time slot and $q$ is the probability that the primary network is in state Vg in the next time slot. This implies that the probability that the primary network is in state B is $(1-p-q)$. After each time slot, depending on the action taken by secondary user and the corresponding feedback, $p$ and $q$ can be updated according to one of the following four cases.

Case 1: Secondary user is silent and a NACK is received from the primary network. The NACK during secondary user silence implies that primary network has been in state B and, thus, the primary receiver has failed to receive the packet. Therefore, the belief state in the next time slot is:
\begin{equation}
p(1)=P_{\rm BG}
\end{equation}
\begin{equation}
q(1)=P_{\rm BVg}
\end{equation}
where, as in the two-state case, $p(k)$ and $q(k)$ are the update expressions for $p$ and $q$, respectively, for the $k$th case.

Case 2: Secondary user is silent and an ACK is received from the primary network. Primary network could be in state G with probability $\frac{p}{p+q}$ or state Vg with probability $\frac{q}{p+q}$. The belief state in the next time slot is:
\begin{equation}
p(2)=\frac{p}{p+q}P_{\rm GG}+\frac{q}{p+q}P_{\rm VgG}
\end{equation}
\begin{equation}
q(2)=\frac{p}{p+q}P_{\rm GVg}+\frac{q}{p+q}P_{\rm VgVg}
\end{equation}

Case 3: Secondary user is transmitting and an ACK is received from the primary network. The ACK during secondary user activity implies that primary network has been in state Vg. Therefore, the belief state in the next time slot is:
\begin{equation}
p(3)=P_{\rm VgG}
\end{equation}
\begin{equation}
q(3)=P_{\rm VgVg}
\end{equation}

Case 4: Secondary user is transmitting and a NACK is received from the primary network. Primary network could be in state G with probability $\frac{p}{1-q}$ or state B with probability $\frac{1-p-q}{1-q}$. The belief state in the next time slot is:
\begin{equation}
p(4)=\frac{p}{1-q}P_{\rm GG}+\frac{1-p-q}{1-q}P_{\rm BG}
\end{equation}
\begin{equation}
q(4)=\frac{p}{1-q}P_{\rm GVg}+\frac{1-p-q}{1-q}P_{\rm BVg}
\end{equation}
Let $Q_i(p,q)$, $i=1,2,3,4$, denotes the probability that case $i$ above happens:
\begin{equation}
Q_1(p,q)=1-p-q\\
\end{equation}
\begin{equation}
Q_2(p,q)=p+q\\
\end{equation}
\begin{equation}
Q_3(p,q)=q\\
\end{equation}
\begin{equation}
Q_4(p,q)=1-q\\
\end{equation}
The parameters $p$ and $q$ characterizing the belief state are updated by one of the previous four conditions.
If secondary user is listening, the expected current gain can be calculated as:
\begin{equation}
G_1(p,q)=wr_{\rm p}(p+q)
\end{equation}
But if the secondary user is transmitting, the expected current gain is:
\begin{equation}
G_2(p,q)=(1-w)r_{\rm s}+wr_{\rm p}q
\end{equation}
The expected current reward is:
\vspace{0.15cm}
\\
$R(k,p,q,a) = \left\{ \begin{array}{rcl}
{G_1(p,q)} & \mbox{if the secondary user is silent in time slot $k$}
 \\ {G_2(p,q)} & \mbox{if the secondary user is active in time slot $k$}
\end{array}\right.$
\vspace{0.15cm}
\\
The optimal strategy is the strategy that maximizes the following discounted reward function
\begin{equation}
E\left\{{\sum^{K+k-1}_{n=k} \alpha^{n-k}*R(n,p_n,q_n,a_n) \hspace{2mm}|\hspace{2mm} p_k=p,q_0=q}\right\}
\end{equation}
$V^K(p,q)$ satisfies the following Bellman equation \cite{Dynamic Programming}:
\begin{equation}
\label{eqn: itrative bellman2}
\begin{split}
V^K(p,q)= {\rm max} \bigg\{ &wr_{\rm p}(p+q)+\alpha \sum^{}_{i=1,2} Q_i(p,q)V^{K-1}(p(i),q(i)), \\
   & (1-w)r_{\rm s}+wr_{\rm p}q+\alpha \sum^{}_{i=3,4} Q_i(p,q)V^{K-1}(p(i),q(i)) \bigg\}
\end{split}
\end{equation}
where
\begin{equation}
\begin{split}
&V^1(p,q)= {\rm max}\left\{wr_{\rm p}(p+q), (1-w)r_{\rm s}+wr_{\rm p}q\right\}
\end{split}
\end{equation}
When $K=\infty$, $V(p,q)$ denote the maximum achievable discounted reward function.
$V(p,q)$ satisfies the following Bellman equation \cite{Dynamic Programming}:
\begin{equation}
\begin{split}
V(p,q)= {\rm max} \bigg\{ &wr_{\rm p}(p+q)+\alpha \sum^{}_{i=1,2} Q_i(p,q)V(p(i),q(i)), \\
   & (1-w)r_{\rm s}+wr_{\rm p}q+\alpha \sum^{}_{i=3,4} Q_i(p,q)V(p(i),q(i)) \bigg\}
\end{split}
\end{equation}
We solve Equation (\ref{eqn: itrative bellman2}) iteratively via approximating the value function at a finite number of belief values on a two dimensional grid representing $p$ and $q$ values. The value function is initialized and then (\ref{eqn: itrative bellman2}) is used to update it. For $p$ and $q$ values not belonging to the grid, interpolation or extrapolation is used. After convergence, the secondary terminal decides whether to transmit or listen based on the term that maximizes $V(p,q)$ at each value of $p$ and $q$.

\section{Multiple Primary Channels}

\label{sec:Multichannel}
In this section, we extend the problem of a single primary channel to multiple independent erasure primary channels with different transition probabilities. The secondary user has to decide which primary channel to access. Here, we focus on the two channel case. However, this technique can be extended to the multi-channel case, though with increasing computational complexity.
\\We have two independent primary channels with two different transition matrices $P_1$ and $P_2$.
\[
P_1 =
\left[ {\begin{array}{cc}
 P_{\rm EE1} & P_{\rm EN1}  \\
 P_{\rm NE1} & P_{\rm NN1}  \\
 \end{array} } \right]
\]
\[
P_2 =
\left[ {\begin{array}{cc}
 P_{\rm EE2} & P_{\rm EN2}  \\
 P_{\rm NE2} & P_{\rm NN2}  \\
 \end{array} } \right]
\]
The system model is exactly similar to the two-state model except that when the cognitive radio is silent, the secondary transmitter can eavesdrop on the ARQ from both primary channels simultaneously. When the secondary user decides to transmit on a specific primary channel, it can make use of the ARQ from the other primary channel but the ARQ from the occupied channel carries no information since simultaneous transmission is assumed to surely result in a NACK.
\\Our objective is to choose the transmission strategy that maximizes the expected value of weighted sum throughput $R$ given by
\begin{equation}
R=\lim_{K \to \infty}\frac{1}{K}\sum_{n=1}^{K}\Bigl(w(R_{\rm p1}(n,p_n,q_n,a_n)+R_{\rm p2}(n,p_n,q_n,a_n))+\left(1-w\right)R_{\rm s}(n,p_n,q_n,a_n)\Bigl)
\end{equation}
where $R_{\rm p1}(n)$ and $R_{\rm p2}(n)$ are the primary throughput of the first and second channels in time slot $n$, respectively, $R_{\rm s}(n)$ is the secondary throughput in time slot $n$ and $0\leq w \leq 1$ is a weighting factor that determines the relative importance of the two rates, $p_n$, $q_n$ are the beliefs of the two primary channels, and $a_n$ is the action taken by the secondary user in time slot $n$.

The cognitive radio action in time slot $n$, $a_n$, is one of the following three different options:
\\1.Secondary user listens to both primary channels.
\\2.Secondary user transmits on the first channel and listens to the second channel.
\\3.Secondary user transmits on the second channel and listens to the first channel.

We can once more apply dynamic programming to this system to find the optimal decisions for the secondary user.
The belief state is parameterized by two parameters $p$ and $q$,
where $p$ is the probability that the first primary network is in erasure state in the next time slot, and $q$ is the probability that the second primary network is in erasure state in the next time slot.
\\After each time slot, depending on the action taken by secondary user and the corresponding feedback, $p$ can be updated according to one of the following 3 cases:

Case 1: secondary user is not transmitting on the first channel and an ACK is received from the first primary network. Therefore, the updated value of $p$ is:
\begin{equation}
p(1)=P_{\rm NE1}
\end{equation}

Case 2: secondary user is not transmitting on the first channel and a NACK is received from the first primary network. Therefore, the updated value of $p$ is:
\begin{equation}
p(2)=P_{\rm EE1}
\end{equation}

Case 3: secondary user is transmitting on the first channel. The probability of erasure at the next time slot $p$ is updated by the Markovian property as follows.
\begin{equation}
p(3)=pP_{\rm EE1}+(1-p)P_{\rm NE1}
\end{equation}
Similarly, the value of $q$ is also updated according to one of the previous 3 cases but for the second primary network:
\begin{equation}
q(1)=P_{\rm NE2}
\end{equation}
\begin{equation}
q(2)=P_{\rm EE2}
\end{equation}
\begin{equation}
q(3)=qP_{\rm EE2}+(1-q)P_{\rm NE2}
\end{equation}
where, $p(k)$ and $q(k)$ are the update expressions for $p$ and $q$, respectively, for the $k$th case.
\\The parameters $p$ and $q$ can be updated by one of the previous conditions according to the secondary user decision and the primary networks outputs.

Now, the secondary user has three possible decisions, whether to transmit on the first channel, to transmit on the second channel or to remain idle and listen to both channels feedback.
\\If secondary user is idle, the expected current gain will be:
\begin{equation}
G_1(p,q)=w(r_{\rm p1}(1-p)+r_{\rm p2}(1-q))
\end{equation}
But if the secondary user transmits on the first channel, the expected current gain is:
\begin{equation}
G_2(p,q)=(1-w)r_{\rm s}+wr_{\rm p2}(1-q)
\end{equation}
In case the secondary user transmits on the second channel, the expected current gain is:
\begin{equation}
G_3(p,q)=(1-w)r_{\rm s}+wr_{\rm p1}(1-p)
\end{equation}
Therefore, the expected current reward is:
\vspace{0.15cm}
\\
$R(k,p,q,a) = \left\{ \begin{array}{rcl}
{G_1(p,q)} & \mbox{if the secondary user is idle in time slot $k$}
 \\ {G_2(p,q)} & \mbox{if the secondary user is transmitting on first channel in time slot $k$}
 \\ {G_3(p,q)} & \mbox{if the secondary user is transmitting on second channel in time slot $k$}
\end{array}\right.$
\vspace{0.15cm}
\\
The optimal strategy is the strategy that maximizes the following discounted reward function
\begin{equation}
E\left\{{\sum^{K+k-1}_{n=k} \alpha^{n-k}*R(n,p_n,q_n,a_n) \hspace{2mm}|\hspace{2mm} p_k=p}\right\}
\end{equation}
$V^K(p,q)$ satisfies the following Bellman equation \cite{Dynamic Programming}:
\begin{equation}
\label{eqn: itrative bellman3}
\begin{split}
V^K(p,q)= {\rm max} \bigg\{ &(w(r_{\rm p1}(1-p)+r_{\rm p2}(1-q))+\alpha (1-p)(1-q)V^{K-1}(p(1),q(1))+ \\
  & \alpha (1-p)qV^{K-1}(p(1),q(2))+\alpha p(1-q)V^{K-1}(p(2),q(1))+  \\
  &  \alpha pqV^{K-1}(p(2),q(2)), (1-w)r_{\rm s}+wr_{\rm p2}(1-q)+\alpha qV^{K-1}(p(3),q(2))+ \\
  &  \alpha (1-q)V^{K-1}(p(3),q(1)), (1-w)r_{\rm s}+wr_{\rm p1}(1-p)+ \\
   & \alpha pV^{K-1}(p(2),q(3))+\alpha (1-p)V^{K-1}(p(1),q(3)) \bigg\}
\end{split}
\end{equation}
where
\begin{equation}
\begin{split}
V^1(p,q)= {\rm max} \{ & w(r_{\rm p1}(1-p)+r_{\rm p2}(1-q)), \\
   & (1-w)r_{\rm s}+wr_{\rm p2}(1-q),~~(1-w)r_{\rm s}+wr_{\rm p1}(1-p) \}
\end{split}
\end{equation}
When $K=\infty$, $V(p,q)$ denote the maximum achievable discounted reward function as before.

We solve Equation (\ref{eqn: itrative bellman3}) iteratively via approximating the value function at a finite number of belief values on a two dimensional grid representing $p$ and $q$ values. The value function is initialized and then (\ref{eqn: itrative bellman3}) is used to update it. For $p$ and $q$ values not belonging to the grid, interpolation or extrapolation is used. After convergence, the secondary terminal decides whether to transmit on the first channel, transmit on the second channel or listen to both channels based on the term that maximizes $V(p,q)$ for each value of $p$ and $q$.

\section{Hidden Markov model (HMM)}

\label{sec:Hidden Markov model}
In previous sections, we assumed that secondary user has full knowledge of the transition probabilities of the primary network system state. Here, we assume that the cognitive user is not aware of these transition probabilities. We now divide the time horizon into two intervals, the first interval is the training period, where the secondary user ``learns'' the transition probabilities. The second interval is the transmission interval where the secondary terminal uses the estimated probabilities in order to derive the optimal transmission policy based on the algorithms presented in the paper so far. We propose using HMM to estimate the probabilities via listening to the feedback channel.

The basic idea behind HMM is that the actual system state is unknown. The only available information is the observation vector, which is a probabilistic function of the state. For example, the three state model can be described as an HMM where the hidden states of the model are B, G and Vg as mentioned in Section \ref{sec:Three-state system model}. The ARQ feedback vector (ACK or NACK) is considered the observation vector. This feedback is generated as mentioned above according to Table III. Now, we introduce the HMM framework to some of the previously mentioned models.

\subsection{Two-State System Model}

This model is exactly the same as the two-state system model discussed in Section \ref{sec:Two-state System model} except that the transition probabilities are unknown. Hence, the secondary transmitter observes the ARQ feedback from the primary network and uses this observation vector to estimate the state transition probabilities of the primary network using an HMM-based scheme. The HMM framework can be applied to the Erasure/Non-Erasure channel model and the Gilbert-Elliot model as well.

Once the transition probabilities are estimated, the same procedure of dynamic programming presented in Section \ref{sec:Two-state System model} is executed to reach the optimal transmission strategy for secondary user to maximize the weighted sum throughput of the primary and secondary networks.

\subsection{Three-State System Model}

HMM framework is also applied to the three-state model, but here, there is a main difference between the three-state model and the two-state model during applying the HMM framework. In the two-state model, the secondary user is always silent during transition probabilities estimation phase because as long as the secondary user is silent, it obtains the maximum knowledge about the state of the primary channel which will help the secondary user estimate the probabilities correctly. On the other hand, in the three state model, the transition probabilities estimation can be refined by both the secondary user silence and transmission. During silence, there is no uncertainty in the detection of the B state because receiving a NACK feedback, indicates directly that the previous state is a B state, but if an ACK feedback is received, there is uncertainty whether the previous state is G or Vg. Also during secondary user transmission, there is no uncertainty in the detection of state Vg. So a learning scheme would have to combine silence and transmission during the training phase. For example, the learning procedure using HMM can start with the secondary user being silent, then these estimation results are considered as an initial condition for the next phase of learning during which the secondary user is transmitting.

\section{Simulation Results}
\label{sec:Simulation Results}
\begin{figure}[!ht]
\centering
  \includegraphics[width=0.5\columnwidth]{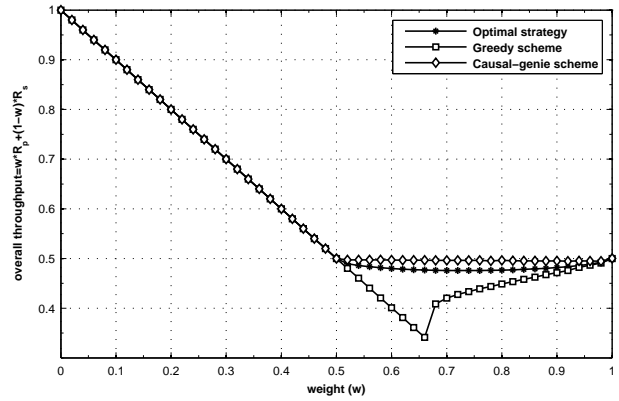}\\
  \caption{Two-state weighted sum throughput.}\label{fig:overall_throughput_2states}
\end{figure}

\begin{figure}[!ht]
\centering
  \includegraphics[width=0.5\columnwidth]{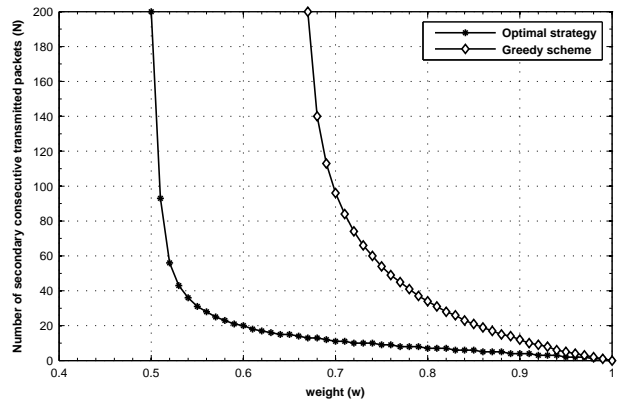}\\
  \caption{Optimal number of consecutive transmitted packets.}\label{fig:M_vs_w}
\end{figure}

\begin{figure}[!ht]
\centering
  \includegraphics[width=0.5\columnwidth]{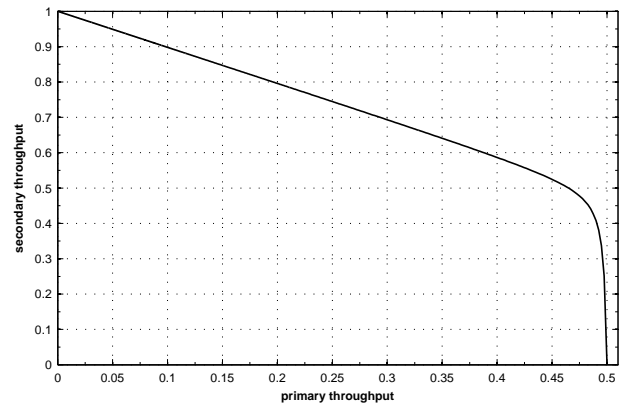}\\
  \caption{primary-secondary rate region.}\label{fig:capacity_region}
\end{figure}

For all our simulation results, the belief is randomly initialized from a uniform distribution, then after simulations, it converges to the actual belief. For the two state erasure/non-erasure channel model, we use the following system parameters:
$P_{\rm EE}=0.99, P_{\rm NE}=0.01, r_p=1$ and $r_s=1$.
The weighted sum of the primary and secondary throughput is shown in Figure \ref{fig:overall_throughput_2states} which shows a significant gain in the throughput for our proposed scheme over the greedy scheme inside the region of interest $w>0.5$. Figure \ref{fig:overall_throughput_2states} also shows that our optimal scheme is very close to the causal-genie scheme. The causal genie scheme is an upper-bound for the performance where we assume that a genie informs the secondary user of the previous primary channel state. Figure \ref{fig:M_vs_w} shows the optimal values of the number of secondary user consecutive transmitted packets $M$ versus different values of the weighting factor $w$. We can see from Figure \ref{fig:M_vs_w} that in the greedy scheme, the secondary transmitter transmits always ($M$ is infinite) as long as $w < 0.67$ which explains the sudden change in the overall throughput at $w = 0.67$ in Figure \ref{fig:overall_throughput_2states}. The optimal strategy has this threshold at $w = 0.5$ which means that the secondary user optimal strategy benefits from learning the channel state rather than transmitting to maximize its immediate reward. Our proposed scheme spans the boundary of the primary-secondary rate region at number of points where $M$ has an integer value. The piecewise linear connection between these points can be achieved by time division multiplexing between different values of integer $M$. For system parameters $r_p=1, r_s=1$ with the same transition probabilities as above, the rate region is shown in Figure \ref{fig:capacity_region}. The rate region is obtained by solving the optimization problem for various values of the weight, $w$.

\begin{figure}[!ht]
\centering
  \includegraphics[width=0.5\columnwidth]{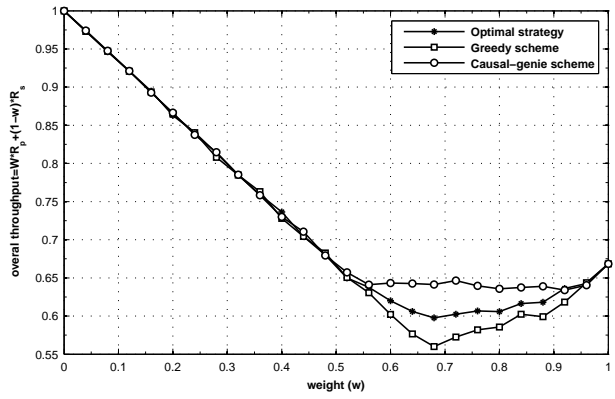}\\
  \caption{Three-state weighted sum throughput.}\label{fig:overall_throughput_3states}
\end{figure}

For the three-state model, the system parameters used for generating the simulation results are as follows:$P_{\rm BB}=P_{\rm GG}=P_{\rm VgVg}=0.9, P_{\rm BG}=P_{\rm GB}=P_{\rm VgG}=0.005, r_p=1$ and $r_s=1$. The weighted sum of the primary and secondary throughput is shown in Figure \ref{fig:overall_throughput_3states}. This again shows that using our scheme of optimization achieves higher throughput than the greedy scheme inside the region of interest $w>0.5$. This again shows that our scheme of optimization with maximizing the total future throughput comes between the greedy scheme of maximizing just the instantaneous reward and the upper-bound of the causal genie.

\begin{figure}[!ht]
\centering
  \includegraphics[width=0.5\columnwidth]{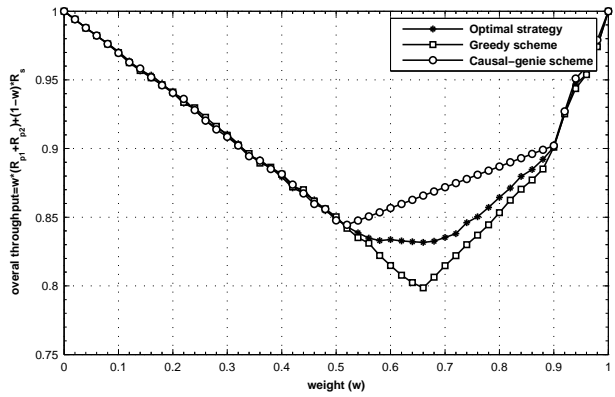}\\
  \caption{Two-channel two-state weighted sum throughput.}\label{fig:two_channel}
\end{figure}

For the case of two primary channels presented in Section \ref{sec:Multichannel}, we show the throughput that we obtain using our optimal transmission strategy compared to the simple greedy model. The two independent primary channels are identical with transition probabilities $P_{EE1}=P_{EE2}=0.99, P_{NE1}=P_{NE2}=0.01$. The secondary user reward is $r_s=1$ and the primary channels rewards are $r_{p1}=r_{p2}=1$. The weighted sum of the primary and secondary throughput is shown in Figure \ref{fig:two_channel}. This figure also shows that our proposed scheme is better than greedy scheme for a region of the weight, $w$, values.

\begin{figure}[!ht]
\centering
  \includegraphics[width=0.5\columnwidth]{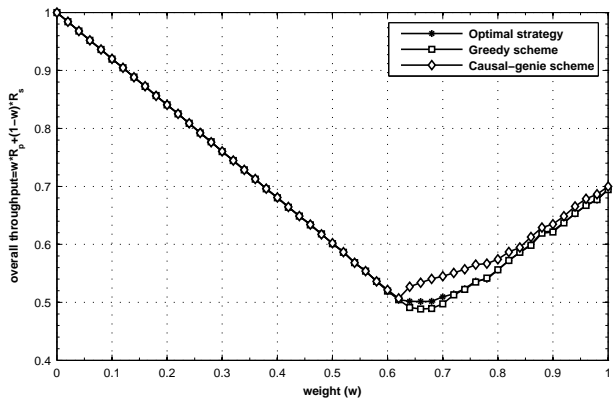}\\
  \caption{Weighted sum throughput for the Gilbert-Elliot Model.}\label{fig:gamma}
\end{figure}

For the Gilbert-Elliot mode, Table I shows the probability of receiving an ACK from the primary network in different channel states and secondary user decisions. The numerical parameters used for the simulation are $\gamma_1=0.2, \gamma_2=0.01, \gamma_3=0.95$ and $\gamma_4=0.3$. The system transition probabilities are $P_{\rm EE}=0.8, P_{\rm NE}=0.1$.Figure \ref{fig:gamma} shows a minor increase of the overall throughput of our proposed scheme over the greedy scheme.

\begin{figure}[!ht]
\centering
  \includegraphics[width=0.5\columnwidth]{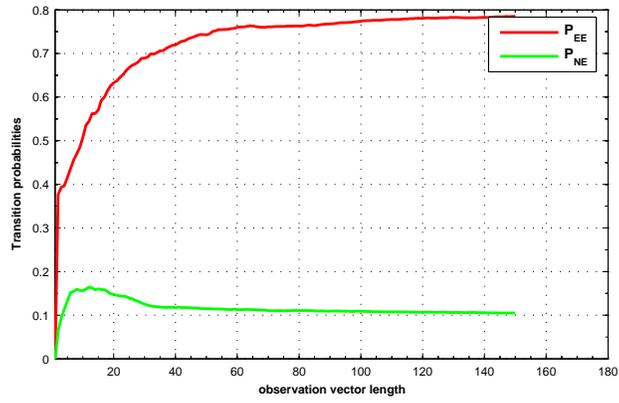}\\
  \caption{Transition probabilities estimation.}\label{fig:estimate_transition_prob}
\end{figure}

\begin{figure}[!ht]
\centering
  \includegraphics[width=0.5\columnwidth]{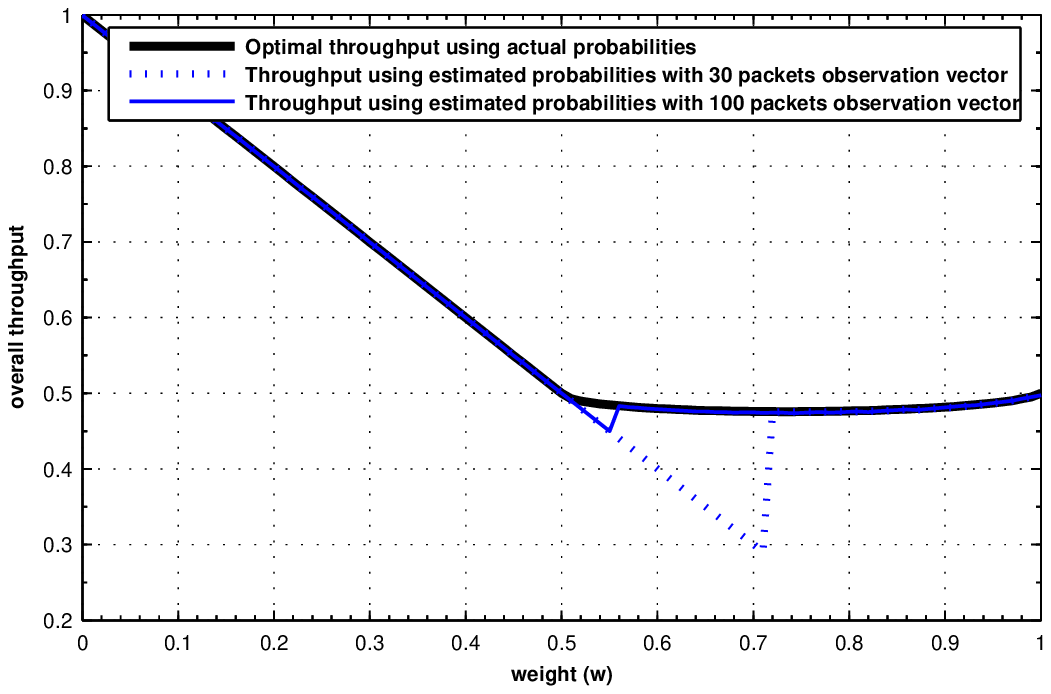}\\
  \caption{Throughput degradation due to estimation error.}\label{fig:throughput_degradation}
\end{figure}

\begin{figure}[!ht]
\centering
  \includegraphics[width=0.5\columnwidth]{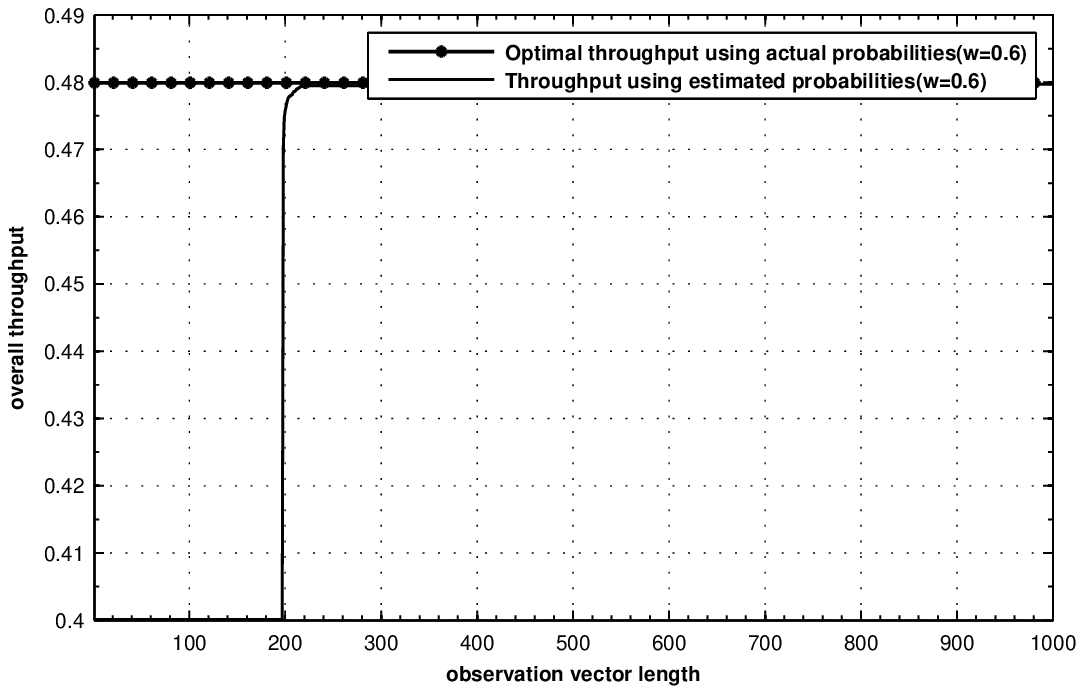}\\
  \caption{Throughput degradation due to estimation error at weight $w=0.6$.}\label{fig:fixed_weight}
\end{figure}

\begin{figure}[!ht]
\centering
  \includegraphics[width=0.5\columnwidth]{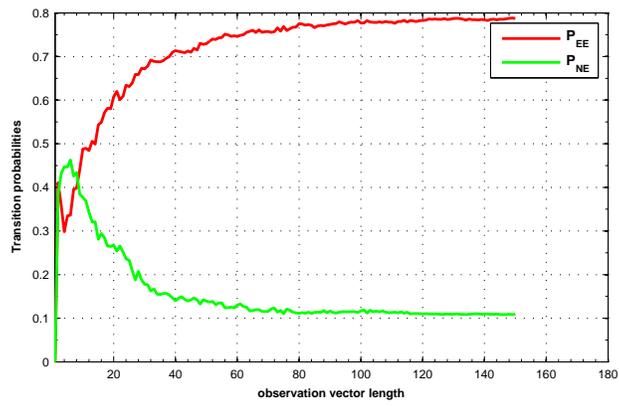}\\
  \caption{Transition probabilities estimation for the Gilbert-Elliot Model.}\label{fig:learning_gamma}
\end{figure}

To evaluate our proposed algorithm of learning the transition probabilities, we simulated the two-state model as follows. For the Erasure/Non-Erasure channel model, we use a sequence of observations representing the ARQ feedback for estimating the primary channel transition probabilities using HMM-based scheme. The actual transition probabilities which need estimation are $P_{\rm EE}=0.99, P_{\rm NE}=0.01$. The estimation of the transition probabilities versus the observation vector length is shown in Figure \ref{fig:estimate_transition_prob}. We can see that estimation is refined by increasing the length of the observation vector. Figure \ref{fig:throughput_degradation} shows the degradation in the overall throughput due to using the estimated probabilities using 30 and 100 feedback packets as an observation vector in the HMM-based scheme instead of the actual probabilities in the optimal strategy calculations. Figure \ref{fig:throughput_degradation} shows how the length of the observation vector can refine the transition probabilities estimation and consequently decrease the throughput degradation. Figure \ref{fig:fixed_weight} shows the throughput degradation as a function of the observation vector length at a fixed weight w=0.6. These results show that we can accurately estimate the transition probabilities and suffer limited loss in throughput if we use around 100 packets in our learning algorithm.
\\For the Gilbert-Elliot model, the actual transition probabilities we estimate are $P_{\rm EE}=0.8, P_{\rm NE}=0.1$. The estimation of the transition probabilities versus the observation vector length is shown in Figure \ref{fig:learning_gamma}. Clearly and expectedly, increasing the length of the observation vector leads to better estimation.
\section{Conclusion}
\label{sec:Conclusion}
In this work, the ACK/NACK feedback from the primary receiver is exploited by the secondary transmitter in order to find optimal transmission strategies that maximize the weighted sum of primary and secondary throughput. We have formulated the problem when the number of primary channel states is ageneral number of states $S$. We then tackled the two-state system model, where we have shown that the optimal transmission strategy is to transmit for a fixed number of packets, $M$, after hearing a NACK. We derived an expression for $M$ and derived a closed-form expression of the optimal overall throughput. For the Gilbert-Elliot model, we have shown that the optimal strategy is a threshold based policy on the belief of the secondary user that the primary link is in the erasure state, and used dynamic programming to obtain the optimal transmission strategy. We have then solved the problem for the case of three channel states and used dynamic programming to obtain the optimal secondary user policy.
For the multiple primary channels model, we derived the optimal strategy for the secondary user to choose the primary channel on which to transmit in order to maximize the current and future reward.
In the case of unknown Markov chain transition probabilities, we proposed using HMM to estimate these probabilities, and have shown via simulations that close to optimal performance can be obtained by using long enough training period.

\section{Appendices}
\subsection{Transmission policy of the two-state system is threshold based}

\label{app:Properties of the Function}

We show in this appendix that the optimal policy of the two-state system (Erasure/Non-Erasure channel model) is threshold based in the belief of the secondary user that the primary channel is in erasure. The reward function $V^K(p)$ is defined as follows.
\begin{equation}
\begin{split}
V^K(p)= {\rm max} \bigg\{ &wr_{\rm p}(1-p)+(1-p)V^{K-1}\left(p(1)\right)+pV^{K-1}\left(p(2)\right), \\
& (1-w)r_{\rm s}+V^{K-1}(p(3)) \bigg\}
\end{split}
\end{equation}
where
\begin{equation}
V^1(p)= {\rm max} \left\{wr_{\rm p}(1-p), (1-w)r_{\rm s}\right\}
\end{equation}
If $wr_{\rm p}<(1-w)r_{\rm s}$, thus, we must have $wr_{\rm p}(1-p)<(1-w)r_{\rm s}$ as $0\leq p \leq1$.
\\This means that
\begin{equation}
V^1(p)= (1-w)r_{\rm s}
\end{equation}
\begin{equation}
\begin{split}
V^2(p)= {\rm max} \bigg\{ &wr_{\rm p}(1-p)+(1-p)V^1\left(P_{\rm NE}\right)+pV^1\left(P_{\rm EE}\right), \\
   & (1-w)r_{\rm s}+V^1(pP_{\rm EE}+(1-p)P_{\rm NE}) \bigg\}
\end{split}
\end{equation}
\begin{equation}
V^2(p)= {\rm max} \left\{wr_{\rm p}(1-p)+(1-w)r_{\rm s}, (1-w)r_{\rm s}+(1-w)r_{\rm s}\right\}
\end{equation}
\begin{equation}
V^2(p)= (1-w)r_{\rm s}
\end{equation}
Thus, if $wr_{\rm p}<(1-w)r_{\rm s}$, we always have
\begin{equation}
V(p)= (1-w)r_{\rm s}
\end{equation}
Hence, secondary user should always transmit.\\
On the other hand, if $wr_{\rm p}>(1-w)r_{\rm s}$
\begin{equation}
V^1(p)= {\rm max} \left\{wr_{\rm p}(1-p), (1-w)r_{\rm s}\right\}
\end{equation}
And, since the maximum of two convex and non-increasing functions is also convex and non-increasing, thus, $V^1(p)$ is convex and non-increasing function in $p$.
\\We now use induction to show that $V^{K}(p)$ is convex and non-increasing for all $K$.
\\Lets assume that $V^{K-1}(p)$ is convex and non-increasing function, we have
\begin{equation}
\begin{split}
V^K(p)= {\rm max} \bigg\{ &wr_{\rm p}(1-p)+(1-p)V^{K-1}\left(P_{\rm NE}\right)+pV^{K-1}\left(P_{\rm EE}\right), \\
   & (1-w)r_{\rm s}+V^{K-1}(p(P_{\rm EE}-P_{\rm NE})+P_{\rm NE}) \bigg\}
\end{split}
\end{equation}
First term is linear in $p$, i.e., it is convex. Second term is also convex in $p$. Hence, by induction, $V^K(p)$ is convex in $p$. Now, since $P_{\rm EE}>P_{\rm NE}$ (assumption), $V^{K-1}(p)$ is non-increasing function in $p$. Thus, $V^{K-1}(P_{\rm EE})<V^{K-1}(P_{\rm NE})$. Thus, first term is non-increasing in $p$ and second term is also non-increasing in $p$. Thus, we can deduce, $V^K(p)$ is non-increasing function in $p$. Hence, $V^K(p)$ is convex and non-increasing function in $p$ for all $K$.
\\At steady state,
\begin{equation}
\label{eqn: steady_state}
\begin{split}
V(p)= {\rm max} \bigg\{ &wr_{\rm p}(1-p)+(1-p)V\left(P_{\rm NE}\right)+pV\left(P_{\rm EE}\right), \\
   & (1-w)r_{\rm s}+V(pP_{\rm EE}+(1-p)P_{\rm NE}) \bigg\}
\end{split}
\end{equation}
Since $V(p)$ is convex
\begin{equation}
V(pP_{\rm EE}+(1-p)P_{\rm NE}) \leq pV\left(P_{\rm EE}\right)+(1-p)V\left(P_{\rm NE}\right)
\end{equation}
Since $(1-w)r_{\rm s} \leq wr_{\rm p}$ (otherwise secondary user is always transmitting)
\begin{equation}
(1-w)r_{\rm s}+V(pP_{\rm EE}+(1-p)P_{\rm NE}) \leq wr_{\rm p}+pV\left(P_{\rm EE}\right)+(1-p)V\left(P_{\rm NE}\right)
\end{equation}
Therefore, the first and second terms of Equation (\ref{eqn: steady_state}) can be written as
\\ 2nd term $\leq$ 1st term $+wr_{\rm p}p$
\\Then, if $p=0$, 2nd term $\leq$ 1st term, thus, the decision is to listen.
\\If $p=1$
\begin{equation}
V(p)= {\rm max} \{ V\left(P_{\rm EE}\right) , (1-w)r_{\rm s}+V(P_{\rm EE}) \}
\end{equation}
$V(p)= (1-w)r_{\rm s}+V(P_{\rm EE})$. Thus, the decision is to transmit.
\\Since the first and second terms of Equation (\ref{eqn: steady_state}) are convex and non-increasing functions in $p$, the optimal strategy is a threshold based strategy in $p$, i.e., there is a unique threshold value of $p$ called $p_{th}$ at which the secondary user converts from listening to transmitting which represent the intersection between first term and second term of Equation (\ref{eqn: steady_state}) plotted versus $p$.
\\Due to convexity of the first and second terms of Equation (\ref{eqn: steady_state}), the intersection $p_{th}$ has an upper and lower bounds as follows.
\begin{equation}
1-\frac{(1-w)r_{\rm s}}{wr_{\rm p}} < p_{th} < 1- \frac{(1-w)r_{\rm s}}{wr_{\rm p}}P_{\rm NE}
\end{equation}
If $P_{\rm EE} < p_{th}$, the secondary user will listen always whether the channel is in erasure or not, and if $P_{\rm NE} > p_{th}$, it will transmit always whether the channel is in erasure or not.
\\Thus, for secondary user finite number of transmission, we must have $P(E) < p_{th} < P_{\rm EE}$ where $P(E)$ is the stationary probabilities of being in erasure as we also mentioned in Section \ref{sec:Two-state System model}
\\i.e., $\frac{P_{\rm NE}}{P_{\rm NE}+P_{\rm EN}} < p_{th} < P_{\rm EE}$.
\subsection{Quasi-concavity of the overall rate}

\label{app:Quasi-concavity}
Here we will show that the expected throughput $R(M)$ (defined in Equation (\ref{eqn: R(M)})) has only a single peak for $M \geq 0$. We will show this via showing that the equation $R(M+1)-R(M)=0$ has only one solution. Using the scheme defined in Section \ref{sec:Two-state System model}, the expected overall throughput $R(M)$ is defined by
\begin{eqnarray}
R(M) & = & wR_{\rm p}+(1-w)R_{\rm s} \nonumber \\
   & = & \frac{wr_{\rm p}P^{ss}_{\rm N}}{P^{ss}_{\rm N}+P^{ss}_{\rm E}+MP^{ss}_{\rm S}}+\frac{(1-w)r_{\rm s}MP^{ss}_{\rm S}}{P^{ss}_{\rm N}+P^{ss}_{\rm E}+MP^{ss}_{\rm S}} \nonumber \\
   & = & \frac{wr_{\rm p}(1-T^M (P_{\rm EE}))+(1-w)r_{\rm s}MP_{\rm NE}}{1-T^M (P_{\rm EE})+P_{\rm NE}+MP_{\rm NE}}
\end{eqnarray}
where
\begin{equation}
T^M (P_{\rm EE})=\frac{P_{\rm NE}+(P_{\rm EE}-P_{\rm NE})^{(M+1)} (1-P_{\rm EE})}{1+P_{\rm NE}-P_{\rm EE}}
\end{equation}
We have
\begin{equation}
R(M)=\frac{wr_{\rm p}(1-P_{\rm EE}-(1-P_{\rm EE})(P_{\rm EE}-P_{\rm NE})^{(M+1)})+(1-w)r_{\rm s}MP_{\rm NE}(1+P_{\rm NE}-P_{\rm EE})}{1-P_{\rm EE}-(1-P_{\rm EE})(P_{\rm EE}-P_{\rm NE})^{(M+1)}+(M+1)P_{\rm NE}(1+P_{\rm NE}-P_{\rm EE})}
\end{equation}
Let
\begin{eqnarray*}
& A & = 1-P_{\rm EE} \nonumber \\
& B & = P_{\rm EE}-P_{\rm NE} \nonumber \\
& P & = P_{\rm NE} \nonumber \\
\end{eqnarray*}
Thus, we have
\begin{equation}
R(M)=\frac{wr_{\rm p}(A-AB^{(M+1)})+(1-w)r_{\rm s}MP(1-B)}{A-AB^{(M+1)}+(M+1)P(1-B)}
\end{equation}
Similarly, we have
\begin{equation}
R(M+1)=\frac{wr_{\rm p}A(1-B^{(M+2)})+(1-w)r_{\rm s}(M+1)P(1-B)}{A(1-B^{(M+2)})+(M+2)P(1-B)}
\end{equation}
Note that for infinite $M$, we have $R(M) \simeq R(M+1)$. For non-infinite $M$, and if there is a peak at a certain value of $M$, we will have a root of $R(M+1)-R(M)=0$. Hence, we want to solve the difference equation $R(M+1)-R(M)=0$ for $M$. By substituting in the rate difference equation and applying some algebraic manipulations, we have
\begin{equation}
\begin{split}
&\left. wr_{\rm p}AP(M+1)(1-B)(1-B^{(M+2)})-wr_{\rm p}AP(M+2)(1-B)(1-B^{(M+1)})+\right. \\
&\left. (1-w)r_{\rm s}(M+1)AP(1-B)(1-B^{(M+1)})-(1-w)r_{\rm s}MAP(1-B)(1-B^{(M+2)})+\right. \\
& \left. (1-w)r_{\rm s}P^2(1-B)^2=0  \right.
\end{split}
\end{equation}
Let
\begin{eqnarray*}
& a & = wr_{\rm p}AP(1-B) \nonumber \\
& b & = (1-w)r_{\rm s}AP(1-B) \nonumber \\
\end{eqnarray*}
Thus, we have
\begin{equation}
\begin{split}
&\left. ((a-b)M+a)(1-B^{(M+2)})-((a-b)M+a)(1-B^{(M+1)})+ \right. \\
&\left. (b-a)(1-B^{(M+1)})=-(1-w)r_{\rm s}P^2(1-B)^2  \right.
\end{split}
\end{equation}
\begin{equation}
((a-b)M+a)(1-B)B^{(M+1)}+(a-b)B^{(M+1)}=(a-b)-(1-w)r_{\rm s}P^2(1-B)^2
\end{equation}
Let
\begin{eqnarray*}
& C & = (a-b)(1-B) \nonumber \\
& D & = (a-b)+a(1-B) \nonumber \\
\end{eqnarray*}
Now, we have
\begin{equation}
B^{(M+1)}(CM+D)=(a-b)-(1-w)r_{\rm s}P^2(1-B)^2
\end{equation}
\begin{equation}
(M+1)\log B+\log(CM+D)=\log((a-b)-(1-w)r_{\rm s}P^2(1-B)^2)
\end{equation}
\begin{equation}
(CM+D)\log B+C\log(CM+D)=C\log((a-b)-(1-w)r_{\rm s}P^2(1-B)^2)+(D-C)\log B
\end{equation}
Let
\begin{equation}
M_1 = CM+D \nonumber
\end{equation}
So, we have
\begin{equation}
M_1\log B+C\log M_1=C\log((a-b)-(1-w)r_{\rm s}P^2(1-B)^2)+(D-C)\log B
\end{equation}
\begin{equation}
\label{eqn: solveM1}
C\log M_1=K+M_1\log B
\end{equation}
where $K=C\log((a-b)-(1-w)r_{\rm s}P^2(1-B)^2)+(D-C)\log B$ is a constant independent of $M_1$.
The solution of the previous equation is the intersection of linear function of $M_1$ and logarithmic function of $M_1$, i.e., the previous equation has at most two solutions.
\\Note that when $wr_{\rm p}<(1-w)r_{\rm s}$, the term $((a-b)-(1-w)r_{\rm s}P^2(1-B)^2)$ is negative, i.e., the secondary user should always transmit which is proved in Appendix \ref{app:Properties of the Function}.
\\Otherwise, when $(1-w)r_{\rm s}<wr_{\rm p}$, we have
\begin{equation}
\label{eqn: log}
\log M_1=\frac{K-M_1\log B}{C}
\end{equation}
Now the roots of the equation $R(M+1)-R(M)$ are the solutions to the above equation, i.e., the intersection between a logarithmic function of $M_1$ and a linear function of $M_1$. Note that both these functions are monotonically increasing $(B<1)$, and since the slope of a logarithm is a decreasing function, then if both a logarithmic function and a linear function intersect at more than one point, then the slope of the logarithmic function has to be higher than the linear function at the first intersection point. Now we show that the maximum slope of the logarithmic function is less than the slope of the linear function.
\\Since $C=(a-b)(1-B)>0$, then the minimum value of $M_1$ is $D$ where $M_1 = CM+D$, i.e., the minimum is achieved when $M=0$. Thus, the maximum slope of the LHS of Equation (\ref{eqn: log}) with respect to $M_1$ is $1/D$ and the slope of RHS is $-\frac{\log B}{C}$.
\\Now we need to show that $\frac{1}{D} < -\frac{\log B}{C}$
\begin{eqnarray*}
& \frac{C}{D} & = \frac{(a-b)(1-B)}{(a-b)+a(1-B)} \nonumber \\
&  & = \frac{(1-\frac{b}{a})(1-B)}{(1-\frac{b}{a})+(1-B)} \nonumber \\
&  & = \frac{(1-\frac{(1-w)r_{\rm s}}{wr_{\rm p}})(1-B)}{(1-\frac{(1-w)r_{\rm s}}{wr_{\rm p}})+(1-B)} \nonumber \\
&  & = \frac{XY}{X+Y}
\end{eqnarray*}
Where $X=1-\frac{(1-w)r_{\rm s}}{wr_{\rm p}}$ and $Y=1-B$.
\\By substitution, we have $-\log B=\log(\frac{1}{1-Y})$.
\\Thus, we have
\begin{equation}
\frac{XY}{X+Y} = \frac{Y}{1+\frac{Y}{X}} < Y < \log(\frac{1}{1-Y})
\end{equation}
Which means that
\begin{equation}
\label{eqn: result}
\frac{1}{D} < -\frac{\log B}{C}
\end{equation}
Equation (\ref{eqn: result}) shows that the maximum slope of the logarithmic term in the LHS of Equation (\ref{eqn: log}) is always less than the slope of the linear term in the RHS of Equation (\ref{eqn: log}) (recall that both the linear and logarithmic functions are monotonic functions).Thus, the linear and logarithmic terms of Equation (\ref{eqn: log}) intersect only once.
\\Thus, the solution of Equation (\ref{eqn: solveM1}) with respect to $M_1$ has a unique solution which leads to the optimal number of secondary user consecutive transmissions $M$.
\subsection{Transmission policy of Gilbert-Elliot model is a threshold based policy}

\label{app:limited feedback}
As we mentioned in Section \ref{sec:limited feedback}, imperfect ACK/NACK reception is assumed. Table I shows the probability of receiving an ACK from the primary network in different channel states and secondary user decisions. The reward function $V^K(p)$ is defined as follows.
\begin{equation}
\begin{split}
V^K(p)= {\rm max} \bigg\{ &wr_{\rm p}P(ACK|idle)+ P(ACK|idle)V^{K-1}(p(4))+ \\
  &  (1-P(ACK|idle))V^{K-1}(p(3)), (1-w)r_{\rm s}+wr_{\rm p}P(ACK|transmit)+ \\
   & P(ACK|transmit)V^{K-1}(p(2)) + (1-P(ACK|transmit))V^{K-1}(p(1)) \bigg\}
\end{split}
\end{equation}
where
\begin{eqnarray*}
& p(1) & = P(E|NACK,transmit)P_{\rm EE}+(1-P(E|NACK,transmit))P_{\rm NE} \nonumber \\
&  & = \frac{(1-\gamma_2)p}{(1-\gamma_2)p+(1-\gamma_4)(1-p)} \big[ P_{\rm EE}-P_{\rm NE} \big]+ P_{\rm NE} \nonumber \\
& p(2) & = P(E|ACK,transmit)P_{\rm EE}+(1-P(E|ACK,transmit))P_{\rm NE} \nonumber \\
&  & = \frac{\gamma_2p}{\gamma_2p+\gamma_4(1-p)} \big[P_{\rm EE}-P_{\rm NE} \big] +P_{\rm NE} \nonumber \\
& p(3) & = P(E|NACK,idle)P_{\rm EE}+(1-P(E|NACK,idle))P_{\rm NE} \nonumber \\
&  & = \frac{(1-\gamma_1)p}{(1-\gamma_1)p+(1-\gamma_3)(1-p)} \big[ P_{\rm EE}-P_{\rm NE} \big]+ P_{\rm NE} \nonumber \\
& p(4) & = P(E|ACK,idle)P_{\rm EE}+(1-P(E|ACK,idle))P_{\rm NE} \nonumber \\
&  & = \frac{\gamma_1p}{\gamma_1p+\gamma_3(1-p)} \big[P_{\rm EE}-P_{\rm NE} \big] +P_{\rm NE} \nonumber \\
\end{eqnarray*}
and
\begin{eqnarray}
 V^1(p) &=& {\rm max} \left\{ wr_{\rm p}(p\gamma_1+(1-p)\gamma_3)~ ,~ (1-w)r_{\rm s}+wr_{\rm p}(p\gamma_2+(1-p)\gamma_4) \right\} \nonumber \\
   &=& {\rm max} \left\{ wr_{\rm p}p(\gamma_1-\gamma_3)+wr_{\rm p}\gamma_3~ ,~ (1-w)r_{\rm s}+wr_{\rm p}p(\gamma_2-\gamma_4)+wr_{\rm p}\gamma_4 \right\} \nonumber
\end{eqnarray}
Let $\gamma_3>\gamma_1$ and $\gamma_4>\gamma_2$ which is an intuitive assumption which indicates that the probability of receiving an ACK in the Non-erasure state is greater than that in the Erasure state. Thus, $V^1(p)$ is convex and non-increasing function in $p$.
\\Similar to the Erasure/Non-Erasure channel model, we use induction to show that $V^{K}(p)$ is convex and non-increasing in its argument and let's assume that $V^{K-1}(p)$ is convex and non-increasing function, we have
\begin{equation}
\label{eqn: app_B1}
\begin{split}
V^K(p)= {\rm max} \bigg\{ & wr_{\rm p}(p\gamma_1+(1-p)\gamma_3)+(p\gamma_1+(1-p)\gamma_3)V^{K-1}(p(4))+\\
  & (1-p\gamma_1-(1-p)\gamma_3)V^{K-1}(p(3)), (1-w)r_{\rm s}+wr_{\rm p}(p\gamma_2+(1-p)\gamma_4)+ \\
   & (p\gamma_2+(1-p)\gamma_4)V^{K-1}(p(2))+(1-p\gamma_2-(1-p)\gamma_4)V^{K-1}(p(1)) \bigg\}
\end{split}
\end{equation}
\begin{eqnarray*}
& \frac{dp(1)}{dp} & = \frac{\big[(1-\gamma_2)p+(1-\gamma_4)(1-p)\big](1-\gamma_2)-(1-\gamma_2)p\big[(1-\gamma_2)
-(1-\gamma_4)\big]}{\big[(1-\gamma_2)p+(1-\gamma_4)(1-p)\big]^2} \big[ P_{\rm EE}-P_{\rm NE} \big] \nonumber \\
&  & = \frac{(1-\gamma_2)(1-\gamma_4)}{\big[(1-\gamma_2)p+(1-\gamma_4)(1-p)\big]^2} \big[ P_{\rm EE}-P_{\rm NE} \big] \nonumber \\
& \frac{dp(2)}{dp} & = \frac{\big[\gamma_2p+\gamma_4(1-p)\big]\gamma_2-\gamma_2p\big[\gamma_2
-\gamma_4\big]}{\big[\gamma_2p+\gamma_4(1-p)\big]^2} \big[ P_{\rm EE}-P_{\rm NE} \big] \nonumber \\
&  & = \frac{\gamma_2\gamma_4}{\big[\gamma_2p+\gamma_4(1-p)\big]^2} \big[P_{\rm EE}-P_{\rm NE} \big] \nonumber \\
\end{eqnarray*}
Similarly,
\begin{eqnarray*}
& \frac{dp(3)}{dp} & = \frac{(1-\gamma_1)(1-\gamma_3)}{\big[(1-\gamma_1)p+(1-\gamma_3)(1-p)\big]^2} \big[ P_{\rm EE}-P_{\rm NE} \big] \nonumber \\
& \frac{dp(4)}{dp} & = \frac{\gamma_1\gamma_3}{\big[\gamma_1p+\gamma_3(1-p)\big]^2} \big[P_{\rm EE}-P_{\rm NE} \big] \nonumber \\
\end{eqnarray*}
Since $P_{\rm EE}>P_{\rm NE}$, we can see that $\frac{dp(1)}{dp}$, $\frac{dp(2)}{dp}$, $\frac{dp(3)}{dp}$, and $\frac{dp(4)}{dp}$ are greater than or equal to zero. This means that $p(1)$, $p(2)$, $p(3)$, and $p(4)$ are non-decreasing in $p$.
\\Lets start with the monotonicity proof of $V^{K}(p)$ in Equation (\ref{eqn: app_B1}).
\\The terms $wr_{\rm p}(p\gamma_1+(1-p)\gamma_3)=wr_{\rm p}(\gamma_3-p[\gamma_3-\gamma_1])$ and $(1-w)r_{\rm s}+wr_{\rm p}(p\gamma_2+(1-p)\gamma_4)$ are non-increasing in $p$ as $\gamma_3>\gamma_1$ and $\gamma_4>\gamma_2$.
\\ Assuming that $V^{K-1}(p)$ is non-increasing in its argument, we study now the monotonicity of $\Omega_1(p)$ where
\begin{equation}
\Omega_1(p)=(p\gamma_1+(1-p)\gamma_3)V^{K-1}(p(4))
+(1-p\gamma_1-(1-p)\gamma_3)V^{K-1}(p(3)) \nonumber
\end{equation}
\begin{equation}
\begin{split}
\frac{d\Omega_1(p)}{dp}= & -(\gamma_3-\gamma_1)V^{K-1}(p(4))
+(p\gamma_1+(1-p)\gamma_3)\frac{dV^{K-1}(p(4))}{dp(4)}\frac{dp(4)}{dp}\\
   & +(\gamma_3-\gamma_1)V^{K-1}(p(3))+(1-p\gamma_1-(1-p)\gamma_3)
   \frac{dV^{K-1}(p(3))}{dp(3)}\frac{dp(3)}{dp} \nonumber
\end{split}
\end{equation}
\begin{equation}
\label{eqn: app_B2}
\begin{split}
\frac{d\Omega_1(p)}{dp}= & (p\gamma_1+(1-p)\gamma_3)\frac{dV^{K-1}(p(4))}{dp(4)}\frac{dp(4)}{dp}
+(1-p\gamma_1-(1-p)\gamma_3)\frac{dV^{K-1}(p(3))}{dp(3)}\frac{dp(3)}{dp}\\
& -(\gamma_3-\gamma_1)[V^{K-1}(p(4))-V^{K-1}(p(3))]
\end{split}
\end{equation}
Note that
\begin{eqnarray*}
& p(3)-p(4) & = \Big[\frac{1-\gamma_1}{(1-\gamma_1)p+(1-\gamma_3)(1-p)}
-\frac{\gamma_1}{\gamma_1p+\gamma_3(1-p)}\Big] p\big[ P_{\rm EE}-P_{\rm NE} \big] \nonumber \\
&  & = \frac{(\gamma_3-\gamma_1)(1-p)p}{[\gamma_1p+\gamma_3(1-p)][(1-\gamma_1)p+(1-\gamma_3)(1-p)]} \big[P_{\rm EE}-P_{\rm NE} \big] \nonumber \\
&  & \geq 0 \nonumber \\
\end{eqnarray*}
i.e., $p(3) \geq p(4)$ and since $V^{K-1}(p)$ is non-increasing in its argument.
Thus, $V^{K-1}(p(4)) \geq V^{K-1}(p(3))$.
\\In Equation (\ref{eqn: app_B2}), since $\frac{dp(3)}{dp}$ and $\frac{dp(4)}{dp}$ are $\geq 0$, and $\frac{dV^{K-1}(p(3))}{dp(3)}$ and $\frac{dV^{K-1}(p(4))}{dp(4)}$ are $\leq 0$. Thus, $\frac{d\Omega_1(p)}{dp} \leq 0$, i.e., $\Omega_1(p)$ is non-increasing in $p$.
\\The same can be shown for $\Omega_2(p)$ where
\begin{equation}
\Omega_2(p)=(p\gamma_2+(1-p)\gamma_4)V^{K-1}(p(2))
   +(1-p\gamma_2-(1-p)\gamma_4)V^{K-1}(p(1)) \nonumber
\end{equation}
by just replacing $\gamma_1$, $\gamma_3$, $p(3)$, and $p(4)$ by $\gamma_2$, $\gamma_4$, $p(1)$, and $p(2)$ respectively.
\\Since $V^{1}(p)$ is a monotonically non-increasing in $p$, by induction $V^{K}(p)$ is non-increasing in $p$.
\\Now, we will proof the convexity of $V^{K}(p)$.
\\The linear terms $wr_{\rm p}(p\gamma_1+(1-p)\gamma_3)$ and $(1-w)r_{\rm s}+wr_{\rm p}(p\gamma_2+(1-p)\gamma_4)$ are convex. Assuming $V^{K-1}(p)$ is convex in $p$,
\begin{equation}
\begin{split}
\frac{d^2\Omega_1(p)}{dp^2}= & 2(\gamma_3-\gamma_1)\frac{dV^{K-1}(p(3))}{dp(3)}\frac{dp(3)}{dp}
-2(\gamma_3-\gamma_1)\frac{dV^{K-1}(p(4))}{dp(4)}\frac{dp(4)}{dp}\\
& +(p\gamma_1+(1-p)\gamma_3)\Big[\frac{d^2V^{K-1}(p(4))}{dp^2(4)}\Big(\frac{dp(4)}{dp}\Big)^2
+ \frac{dV^{K-1}(p(4))}{dp(4)}\frac{d^2p(4)}{dp^2} \Big]\\
& +(1-p\gamma_1-(1-p)\gamma_3)\Big[\frac{d^2V^{K-1}(p(3))}{dp^2(3)}\Big(\frac{dp(3)}{dp}\Big)^2
+ \frac{dV^{K-1}(p(3))}{dp(3)}\frac{d^2p(3)}{dp^2} \Big]
\end{split}
\end{equation}
where
\begin{eqnarray*}
& \frac{d^2p(3)}{dp^2} & = -\frac{2(1-\gamma_1)(1-\gamma_3)(\gamma_3-\gamma_1)}{\big[(1-\gamma_1)p+(1-\gamma_3)(1-p)\big]^3} \big[ P_{\rm EE}-P_{\rm NE} \big] \nonumber \\
& \frac{d^2p(4)}{dp^2} & = \frac{2\gamma_1\gamma_3(\gamma_3-\gamma_1)}{\big[\gamma_1p+\gamma_3(1-p)\big]^3} \big[P_{\rm EE}-P_{\rm NE} \big] \nonumber \\
\end{eqnarray*}
\begin{equation}
\begin{split}
\frac{d^2\Omega_1(p)}{dp^2}= & (p\gamma_1+(1-p)\gamma_3)\frac{d^2V^{K-1}(p(4))}{dp^2(4)}\Big
(\frac{dp(4)}{dp}\Big)^2+ \\
& (1-p\gamma_1-(1-p)\gamma_3)\frac{d^2V^{K-1}(p(3))}{dp^2(3)}
\Big(\frac{dp(3)}{dp}\Big)^2
\end{split}
\end{equation}
since $\frac{d^2V^{K-1}(p(3))}{dp^2(3)}$ and $\frac{d^2V^{K-1}(p(4))}{dp^2(4)}$ are $\geq 0$.
\\Therefore, $\frac{d^2\Omega_1(p)}{dp^2} \geq 0$. The same procedure can be applied to prove that $\frac{d^2\Omega_2(p)}{dp^2} \geq 0$.
\\By induction, $V^{K}(p)$ is convex.
\\The two terms of $V^{K}(p)$ in Equation (\ref{eqn: app_B1}) are convex and non-increasing. Thus, the optimal strategy is a threshold based policy in $p$ and there is a unique optimal solution.

\section*{Figures}

\end{document}